\documentclass{article}
\usepackage{graphicx}
\usepackage{gensymb}
\usepackage{overpic}
\usepackage{natbib}
\usepackage{subfig}
\graphicspath{{converted_graphics/}}
\usepackage[T1]{fontenc}
\usepackage{lineno}
\usepackage{setspace}
\usepackage{amssymb}
\usepackage{amsmath}
\usepackage{authblk}
\usepackage[margin=1in]{geometry}
\usepackage{color}
\usepackage{hyperref}

\title{The evolution of forecasting \\
for decision making in dynamic environments}

\author[a*]{Andrew R. Tilman}
\author[bc]{V\'itor V. Vasconcelos}
\author[d]{Erol Ak\c{c}ay}
\author[d]{Joshua B. Plotkin}

\begin{document}
\affil[a]{USDA Forest Service, Northern Research Station, St. Paul, MN, USA}
\affil[b]{Informatics Institute, University of Amsterdam, Amsterdam, The Netherlands}
\affil[c]{Institute for Advanced Study, University of Amsterdam, Amsterdam, The Netherlands}
\affil[d]{Department of Biology, University of Pennsylvania, Philadelphia, PA, USA}
\affil[*]{Corresponding Author, Email: andrew.tilman@usda.gov}
\maketitle
\onehalfspacing

\begin{abstract}
\noindent Global change is reshaping ecosystems and societies. Strategic choices that were best yesterday may be sub-optimal tomorrow; and environmental conditions that were once taken for granted may soon cease to exist. In this setting, how people choose behavioral strategies has important consequences for environmental dynamics. Economic and evolutionary theories make similar predictions for strategic behavior in a static environment, even though one approach assumes perfect rationality and the other assumes no cognition whatsoever; but predictions differ in a dynamic environment. Here we explore a middle ground between economic rationality and evolutionary myopia. Starting from a population of myopic agents, we study the emergence of a new type that forms environmental forecasts when making strategic decisions. We show that forecasting types can have an advantage in changing environments, even when the act of forecasting is costly. Forecasting types can invade but not overtake the population, producing a stable coexistence with myopic types. Moreover, forecasters provide a public good by reducing the amplitude of environmental oscillations and increasing mean payoff to forecasting and myopic types alike. We interpret our results for understanding the evolution of different modes of decision-making. And we discuss implications for the management of environmental systems of great societal importance.
\end{abstract}

\section*{Introduction}

Social-ecological systems are characterized by feedback that links ecological states and social decisions. These linkages can induce complex and cyclical dynamics in both human behavior and ecological states~\citep{levin2013social,bieg2017,tilman2018revenue}, driving the dynamics in systems of great societal importance. Models of social-ecological 
systems allow us to identify the key features that determine qualitative outcomes~\citep{schlueter2012new}, and help us to guide management~\citep{farahbakhsh2022}. 

The long-term outcome of a social-ecological system depends on how humans make behavioral decisions, and update their strategies over time. Do individuals deliberate and reason about their alternatives, or do they follow rules of thumb and make rapid, automatic decisions~\citep{kahneman2003,evans2008,kahneman2011thinking}? Dual-process theories of cognition assert that individuals make decisions with both fast and slow thinking, depending upon the context~\citep{pennycook2017}. Thinking fast has been associated with making decisions based on immediate rewards and costs or by intuition, whereas slow, deliberative decision-making is  associated with weighing longer-term implications~\citep{mcclure2004,rand2016}. While dual-process theory is typically studied within an individual, \cite{tomlin2015} found significant population-level consequences of dual-process cognition in human-environmental systems. And the interplay between individuals who adopt automatic versus controlled decision-making is known to drive cyclical dynamics~\citep{rand2017}.

The emerging field of eco-evolutionary games provides a natural tool to model cognition in social-ecological systems \citep{weitz,estrela2019,tilman2020,wang2020steering,wang2020eco,lin2019spatial,barfuss2020caring}.  Eco-evolutionary games arise when the strategies adopted by individuals influence the state of the environment, and the environment in turn influences game payoffs and strategic behavior, generating feedback. There is therefore a close relationship between models of social-ecological systems and the theory of eco-evolutionary games. In both cases, even simple systems can generate persistent oscillations with an ever-changing environment~\citep{weitz,bieg2017,tilman2020}.

Climate change is the result of decades of individual, corporate, and national decisions that have applied strong forcing to our climate system, such that an overshoot of the  2\degree C  warming target is now  likely~\citep{ipcc}. This overshoot is characteristic of the environmental oscillations seen in the theory of eco-evolutionary games~\citep{menard2020conflicts,tilman2020}. The slow timescale at which the environment responds to our strategic behaviors sets the stage for this overshoot. Understanding this system as an eco-evolutionary game may provide important lessons for management and mitigation.

Cyclic dynamics also occur in fisheries, where harvester-driven collapse is ubiquitous~\citep{pauly1998, essington2015}, which leads to reduced profitability of harvesting and sometimes eventual recovery of fish stocks~\citep{hutchings2000, worm2009}. Understanding how modes of decision-making can aggravate or alleviate these human-driven environmental changes has profound practical implications. When should managers expect human behavior and decision-making to produce stable outcomes, as opposed to an unstable or oscillating environment~\citep{bieg2017}?

Forest management can also be viewed as a social-ecological system that undergoes cyclic dynamics of fire, harvesting and regrowth~\citep{luce2012climate,steelman2016}. In the US, a decades-long management emphasis on fire suppression led to increased tree density in western forests~\citep{fellows2008}, which then contributed to increasing wildfire risk~\citep{marlon2012}. This growing risk has been compounded by climate change, which manifests as increased drought frequency and severity, and greater peak summer temperatures~\citep{mckenzie2004climatic}. The USDA Forest Service has developed a strategy for confronting the wildfire crisis which relies heavily on controlled burns and other fuels treatments~\citep{wildfirecrisis2022}. However, shifting from a strategy of fire suppression to a strategy of deploying fire to generate fire-resilient landscapes poses major challenges given the current high fuel loads, climatic conditions, and increased extent of the wildland-urban interface~\citep{radeloff2018rapid}. While the local patterns of wildfire are inherently stochastic, there may be predictable relationships at the landscape and regional scales between management strategies, forest ecosystem states, and wildfire risks. Here again, eco-evolutionary games can provide a useful tool to model these interactions and inform management.    

More generally, long term outcomes in diverse social-ecological systems depend on how individuals make decisions. The psychology of decision-making integrates experiences from the past, observations about the present, and expectations for the future ~\citep{zimbardo1999}. Time discounting measures individuals' inter-temporal preferences for rewards and costs, and it has been measured empirically in humans and other animals~\citep{mischel1989, odum2011}. Different discount rates imply different degrees of future orientation in thinking and decision-making. These differences in the weight placed on expectations for the future is known to impact environmental behaviors~\citep{carmi2013,carmi2014,enzler2019}. Theoretical work has also highlighted the impact of foresight on cooperation~\citep{perry2020foresight}. And, yet, much of the work on social-ecological modeling has assumed strategies and behaviors based in the current payoffs alone, disregarding projections for the future. In this paper, we will explore how individuals who forecast future environmental states might emerge in a population of myopic decision makers, and how forecasters then alter environmental dynamics.

Economic and evolutionary theory both address the problem of decision-making in strategic settings. But economic and evolutionary analyses make vastly different assumptions about what information is available to individuals and how individuals use that information. In classical economic analyses, equilibrium strategies are identified such that fully informed and perfectly rational agents have no incentive to deviate unilaterally -- that is, a Nash equilibrium~\citep{nash1950}. On the other hand, evolutionary game theory does not require individuals to know much, if anything, about the game they are playing~\citep{maynardsmith} but, instead, considers individuals who make strategic decisions through a myopic search process. Such agents that lack cognition altogether nonetheless typically approach an equilibrium called an evolutionary stable strategy (ESS)~\citep{maynardsmith}. Remarkably, in static environments, these two concepts are tightly coupled: all ESS's are Nash equilibria, and all strict Nash equilibria are ESS's. 

The remarkable concordance between the long-term behavior of fully rational individuals versus the behavior of simple myopic agents does not carry over when strategic interactions take place in a changing environment, or when the strategies themselves generate environmental feedback. In a changing environment, a strategy that is favorable today may be detrimental tomorrow. In this setting, there is a wide gulf between the behaviors predicted by traditional game theory and bioeconomic theory~\citep{clark} versus those that arise dynamically in an evolving population of myopic agents. 

Here, we explore a middle ground between myopic agents and fully rational agents -- that is, between evolutionary and economic theories of decision-making. The goal is not to determine which models of decision-making or strategic behavior are correct. Rather, we take the myopic setting as a starting point and ask when individuals will emerge who use more information about their environment and the game they are playing in making strategic decisions. That is, we study how more sophisticated modes of decision-making can arise in a population and what effects this has on environmental dynamics. In particular, we study the emergence of decision-making by individuals who forecast the future state of the environment and account for future payoffs based on their forecasts. Such individuals are not strictly myopic but also not fully rational or perfectly informed. 

Our primary goal is to understand both the emergence of forecasting types as well as their resulting impact on strategic and environmental dynamics. Whereas a population composed entirely of myopic individuals can experience persistent cycles, we will show that one composed entirely of forecasters can produce a stable equilibrium. Moreover, the average fitness in the population of forecasters can be higher than it would be in the myopic population, even when forecasters pay a fixed cost, representing the cognitive or economic burden of making environmental forecasts. We will show that forecasting types can invade a myopic population, but their invasion is self-limiting. Once forecasters stabilize the environment, then forecasting no longer has value (the future will be just like the present), and myopic individuals outperform the forecasters by avoiding the cost of forecasting. As a result, forecasters tend to increase when rare, dampening environmental cycles, but they seldom overtake a population altogether.

We will also show that forecasters and myopic agents can stably persist together and that this coexistence generates both public and private benefits. First, forecasting generates public benefit because even a small sub-population of forecasters serves to reduce the magnitude of environmental variability, which increases the average fitness of both forecasting and myopic types when they coexist. Second, forecasting also generates private benefits because forecasters anticipate environmental change and are better able to deploy the right strategies at the right times. The private benefits allow the evolutionary emergence of forecasters, who then provide public benefits to all.

\section*{Model}
A broad class of eco-evolutionary games produce cyclic dynamics ~\citep{weitz,tilman2020}. These cycles are driven by the strategic decisions of myopic agents. Taking this setting as a starting point, we model the emergence of decision makers who use environmental forecasting in a resident population of myopic agents. Myopic individuals update their strategy following standard replicator dynamics. They switch between strategies based on the current payoffs they experience. Their choices feedback to alter the state of the environment. But myopic individuals do not know anything about the underlying environment; they attend only to the strategies that others employ and the resulting payoffs they receive. 

In contrast to myopic types, forecasting types know and collect more information when making their strategic choices. Forecasting individuals make forecasts of the future states of the environment, and they account for how the changing environment will influence their future payoffs, discounting the expected future payoffs in combination with present value of each alternative strategy. We also model this with replicator dynamics. 

To model the emergence of forecasting, we must also consider the process by which forecasting and myopic types compete. We assume that the act of forecasting carries a cognitive or economic cost. While forecasters may believe that their current strategy will yield large payoffs in the future, this does not give them an advantage over myopic types in the present. Thus, we assume that forecasting and myopic types compete with each other based solely on their instantaneous fitness.  

With these assumptions, the deck is stacked against forecasters. How could a population of forecasters emerge when each forecaster is at an inherent fitness disadvantage and must compete with myopic types based only on instantaneous payoffs? We find that, nonetheless, environmental forecasting allows individuals to anticipate when near-term environmental changes will cause a different strategy to be favored, leading to instantaneous advantages that allow forecasters to invade.

\subsubsection*{Eco-evolutionary games}
We assume that agents are engaged in a two-strategy eco-evolutionary game, building on the  framework developed by ~\cite{tilman2020}. The two strategies are called the ``low-impact" $L$ and ``high-impact" $H$ alternatives, to denote the magnitude of their effects on the environmental state. When all individuals follow the low-impact strategy, the environmental state tends towards its highest value. But when the high-impact strategy dominates in the population, the state of the environment declines towards its lowest possible value. The environmental state is described by a normalized quantity, $n$, that is bounded between zero and one. The payoffs for the eco-evolutionary game are assumed to be linear in the state of the environment, $n$, and in the frequencies of the low- and high-impact strategies. Therefore, the game can be represented by an environmentally dependent payoff matrix:

\begin{equation}
    \Pi(n)=(1-n)
    \begin{bmatrix}
R_0&S_0\\
T_0&P_0
\end{bmatrix}
+n\begin{bmatrix}
R_1&S_1\\
T_1&P_1
\end{bmatrix},
\end{equation}
where the payoff to the low-impact strategy corresponds to the first row of the matrix, and the payoff to the high-impact strategy corresponds to the second row of the matrix. 

Allowing for both forecasting and myopic types of agents, there are in total four kinds of individuals in the population: forecasting or myopic types that follow either strategy $L$ or strategy $H$. We use $z_L^m$ to denote the frequency of $L$-strategy myopic individuals, $z_H^m$ the frequency of $H$-strategy myopic individuals, $z_L^f$  the frequency of $L$-strategy forecasting individuals, and $z_H^f$  the frequency of $H$-strategy forecasting individuals. The overall frequency of strategy $L$ in the whole population is then $z_L=z_L^f+z_L^m$. And the overall frequency of forecasting types is $z^f=z_L^f+z_H^f$.
\begin{figure}
    \centering
    \includegraphics[width = .75\textwidth]{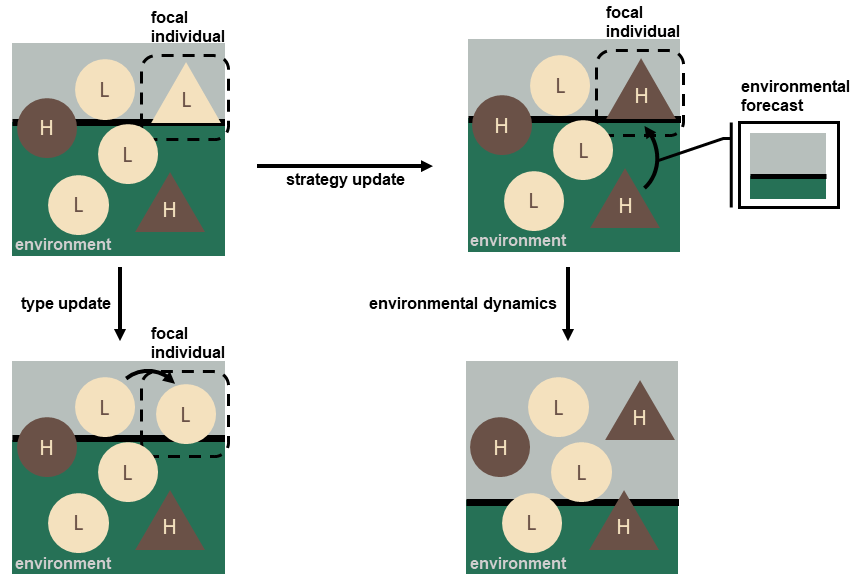}
    \caption{Individuals interact in a changing environment, represented here by the level of the environmental resource shown in green. At any point in time, a focal individual may update their strategy or their type. When a forecasting type (represented by a triangle) updates their strategy, they account for expected future environmental changes. A focal individual may also switch types, from a forecaster to a myopic agent, or vica versa. The environment responds to the frequency of high- and low-impact strategies (shown in brown and cream, respectively) in the population. An individual's type has no direct impact on environmental dynamics, but forecasting and myopic types often adopt different strategies and so environmental dynamics respond indirectly to the composition of forecasting and myopic types in the population. We study the emergence of decision-makers who use forecasting, and what effects they have on environmental dynamics.
    }
    \label{process}
\end{figure}

The immediate payoffs to each strategy are 

\begin{align}
        \pi_L(n,z_L)= (1-n)(R_0 z_L +S_0 (1-z_L)) + n(R_1 z_L + S_1 (1-z_L)), \\
        \pi_H(n,z_L)= (1-n)(T_0 z_L +P_0 (1-z_L)) + n(T_1 z_L + P_1 (1-z_L)). 
\end{align}
Myopic types consider only these immediate payoffs when updating their strategies. In contrast, forecasting types also account for their expectations of the future. These expectations integrate both environmental forecasting and payoff discounting.

\subsubsection*{Forecasting and Discounting}
When forecasting the future state of the environment, forecasting types form expectations about future payoffs.  We assume that forecasters observe the current rate of environmental change and use linear extrapolation to project future environmental states, 

\begin{equation}
    \hat{n}(t)=n+t\dot{n}
\end{equation}
where $\hat{n}(t)$ is the forecasted state of the environment $t$ units of time into the future, $n$ is the current state of the environment, and $\dot{n}$ is the current rate of change of the environment. This linear forecast will be accurate in the short term, but it becomes increasingly unreliable over the long term.  Forecasts are continuously updated, so the accuracy of near-term predictions will remain high throughout the process.

A forecasting type must place some value on their future payoffs for their forecast to have any effect on their behavior. We assume that forecasting individuals put greater weight on their near-term payoffs than long-term payoffs. We model this through a normalized discount function of the form

\begin{equation}
    \omega(t)=re^{-rt}
\end{equation}
where $\omega(t)$ is the weight given to time $t$ into the future and $r$ is the discount rate. This discounting function is normalized so that regardless of $r$, $\omega(t)$ integrates to $1$. Thus, increasing $r$ shifts an agent's preferences toward the near term without changing the total weight that they give to the expected payoffs from the game being played.
\begin{figure}
    \centering
    \includegraphics[width = \textwidth]{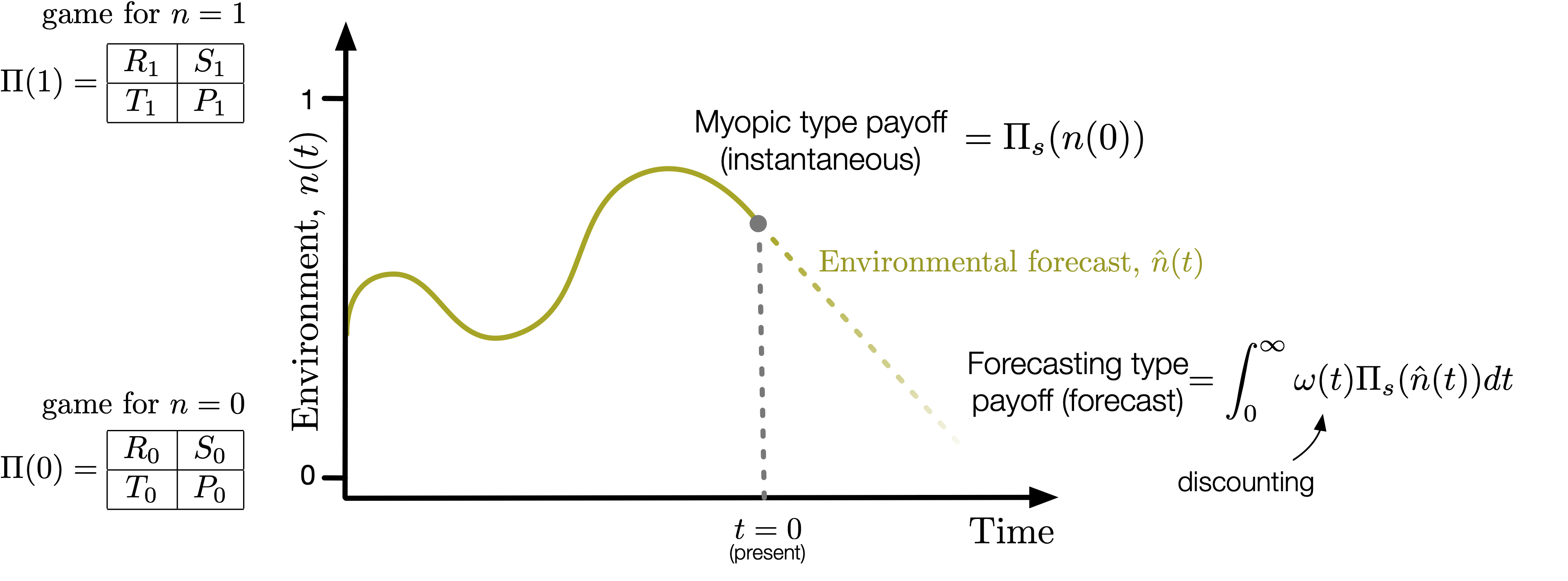}
    \caption{We study decision-making and the emergence of cognitive complexity in a dynamic environment. The game payoff matrix $\Pi$ for two alternative strategies depends on the state of the environment, $n(t)$, which changes over time in response to individuals' actions. A myopic type individual chooses between two alternative strategies based only on the instantaneous payoff of each strategy $s$ in the current environment, $\Pi_s(n(0))$. Whereas a more complex type forms a forecast about the future environmental states, $\hat{n}(t)$ using linear extrapolation, and makes decisions that account for expected future payoffs, discounting the future relative to the present. Starting from a population of purely myopic agents, we show that evolution can favor the emergence of forecasting types, even when they pay a cost to produce forecasts.}
    \label{conceptual}
\end{figure}

\subsubsection*{Eco-evolutionary dynamics}
We model strategic and environmental change based on pairwise interactions among the four possible kinds of individuals in the population. We assume that the act of forecasting is inherently costly. 

For strategy change among forecasting types, discounting and forecasting are integrated into their assessment of the expected payoffs to the low- and high-impact strategies. This leads to the expected payoff of the low-impact strategy,

\begin{equation}
    f_L = \int_0^\infty \omega(t)\pi_L(\hat{n}(t),z_L) \mathrm{d}t
\end{equation}
which can be solved exactly as

\begin{align}
     f_L(z_L,n,\dot{n})&=
     \pi_L(n,z_L)+\frac{\dot{n}}{r}\frac{\partial\pi_L}{\partial n}(n,z_L).
\end{align}
An equivalent expression holds for the high-impact strategy.

For myopic individuals the immediate payoff difference drives strategy dynamics, that is

\begin{equation}
   \pi_L(n,z_L)-\pi_H(n,z_L).
\end{equation}
For forecasting individuals, the payoff difference between the strategies is

\begin{equation}
    f_L-f_H = \pi_L(n,z_L)-\pi_H(n,z_L) + \frac{\dot{n}}{r}\frac{\partial }{\partial n}\left[\pi_L(n,z_L)-\pi_H(n,z_L)\right],
    \label{forecastSelection}
\end{equation}
where a small value of the discount rate $r$ means that agents put a larger weight on future payoffs. Equation~\ref{forecastSelection} shows how forecasting and discounting alter the perception of fitness differences between strategies for forecasters, relative to myopic types. In particular, the first term in Equation ~\ref{forecastSelection} is the immediate difference in payoff between the $L$ and $H$ strategies -- which is the quantity that alone determines myopic decision-making --  whereas the second term reflects how forecasting changes the decision-making process. Note that in the special case that the environment is not changing, that is $\dot{n}=0$, then the payoff difference perceived by forecasters equals the payoff difference perceived by myopic types-- which makes sense because, in this case, the forecasted environment is constant and so forecasting and myopic types make the same strategic decisions.

Now we develop dynamic equations that describe how the frequencies of all four kinds of individuals change in the population. First, we consider $L$-strategy myopic individuals. These individuals can emulate $H$-strategy myopic individuals, or they can choose to adopt either strategy of forecasting individuals. This leads to three terms in the rate of change of $z_L^m$,  following mass-action:

\begin{equation}
    \dot{z}_L^m = z_L^m z_H^m \left(\pi_L-\pi_H\right) + \epsilon_2 z_L^m z_L^f C + \epsilon_2 z_L^m z_H^f \left(\pi_L-\pi_H+C\right).
\end{equation}
The first term corresponds to replicator dynamics among myopic individuals. The second term corresponds to the flux of $L$-strategy forecasters into the myopic $L$-strategists caused by the cost of forecasting: all $L$-strategists receive the same immediate payoff from the game at any point in time, but those who are forecasters pay a fixed cost $C$ for the forecasting ability. Herein lies the dilemma of forecasting. 

The last term represents the flux  of $H$-strategy forecasters to $L$-strategy myopic individuals, driven both by the (immediate) payoff advantage of playing strategy $L$ $\left(\pi_L-\pi_H\right)$ and by the cost of forecasting ($C$). The timescale of strategy switching is likely to be faster than the timescale of type switching. The parameter $\epsilon_2$ controls the relative timescale of type switching as compared to strategy switching.

Next, we consider $L$-strategy forecasting individuals. The dynamics of this sub-population of types follow

\begin{equation}
  \dot{z}_L^f =  z_L^f z_H^f (f_L-f_H) - \epsilon_2 z_L^f z_L^m C + \epsilon_2 z_L^f z_H^m \left(\pi_L-\pi_H-C\right).
\end{equation}
The first term reflects that forecasters switch between strategies $L$ and $H$ based on their perception of the net present value of each strategy, given forecasting and discounting. The second term reflects that forecasters are never favored over myopic types within strategy $L$ or $H$, because they pay a cost to forecast, but the rate of switching types (myopic or forecasting) may be slower than the rate of switching strategies ($L$ or $H$) if $\epsilon_2<1$. The last term shows that $L$-strategy forecasters transition to $H$-strategy myopic individuals according to the immediate payoff difference between the strategies and the cost of forecasting.

We can construct the other equations, for $\dot{z}_H^f$ and $\dot{z}_H^m$ in a similar manner, leading to a dynamical system governed by four equations for the four types/strategies, plus one equation for the linked environmental dynamic: 

\begin{align}
\dot{z}_L^m &= z_L^m z_H^m (\pi_L-\pi_H) + \epsilon_2 z_L^m z_L^f C + \epsilon_2 z_L^m z_H^f (\pi_L-\pi_H+C)\\
\dot{z}_H^m &= - z_H^m z_L^m (\pi_L-\pi_H) + \epsilon_2 z_H^m z_L^f (-\pi_L+\pi_H+C) + \epsilon_2 z_H^m z_H^f C\\
\dot{z}_L^f &= z_L^f z_H^f (f_L-f_H) - \epsilon_2 z_L^f z_L^m C + \epsilon_2 z_L^f z_H^m (\pi_L-\pi_H-C)\\
\dot{z}_H^f &= -z_H^f z_L^f (f_L-f_H) + \epsilon_2 z_H^f z_L^m (-\pi_L+\pi_H-C) - \epsilon_2 z_H^f z_H^m C\\
\dot{n}_{~} &= \epsilon_1(z_L^f+z_L^m-n).
\end{align}
The equation for the environmental dynamic describes a decaying environmental variable \citep{tilman2020}, which can correspond to, for example, pollution levels. In this case, the high- and low-impact strategies generate emissions of the resource. \citet{tilman2020} show that systems with either decaying or renewing intrinsic environmental dynamics (corresponding, for example, a harvested population) generate qualitatively similar eco-evolutionary game dynamics. The timescale of environmental dynamics can differ from the timescales of strategy and type dynamics. Here, $\epsilon_1$ is the relative timescale of environmental dynamics, compared  to strategy dynamics. In total, there are three timescales in the model. Without loss of generality, we consider relative timescales, with strategy dynamics as the reference point. Thus there are only two timescale parameters in the model.

This formulation of the eco-evolutionary game rests on the assumption of mass-action kinetics, where there must be an encounter between any pair of types in order for a transition to occur between them. The likelihood of encounters is proportional to relative abundance, and so encounters among forecasters with different strategies will be rare when forecasting types are rare overall. We focus on the case where type switching is slow relative to strategy switching ($\epsilon_2<1$). We also assume that transitions between forecasting types and myopic types are based on the immediate payoff difference, not based on forecasters' expectations of net present payoffs. In Supplementary Information Section~\ref{IBM} we present an individual-based micro-level model that converges to the system we study in the large population, weak selection limit. 

An alternative formulation of competitive dynamics would consider a hierarchical form of strategy and type imitation, where strategy dynamics within each type are unaffected by the overall abundance of the type (forecaster or myopic) and follow standard two-strategy replicator dynamics. In this formulation, switching between types is based on the mean fitness difference between forecasting and myopic types. When a myopic individual becomes a forecaster (and vice versa), they  choose a strategy (L or H) in proportion to the current frequency of each strategy within that type. We analyze this alternative formulation in Supplementary Information Section~\ref{altDynam}, whereas we focus on the mass-action model across all four strategies/types in the main text.

\section*{Results}

\subsubsection*{Myopic types alone}
We focus our analysis on eco-evolutionary games in which a population of myopic individuals generates cyclical dynamics. These parameter regimes have been identified by prior work on eco-evolutionary games with myopic agents~\citep{tilman2020}. Figure~\ref{temporal} panel (a) shows the temporal dynamics in such a population of myopic individuals, which approaches a limit cycle where the state of the environment and the frequencies of high- and low-environmental impact strategies oscillate. These persistent cycles reduce the long-term average fitness of the population compared to the fitness that could be achieved under a stable environment.

\subsubsection*{Forecasting types alone}
In contrast to a myopic population, a population consisting entirely of forecasters generates a qualitatively different outcome. Provided forecasters care sufficiently about the future (i.e., have a sufficiently small discount rate), then forecasting types can produce a stable outcome for the same eco-evolutionary game that would exhibit cyclical dynamics in a myopic population. Figure~\ref{temporal} panel (b) illustrates a case where a population of forecasters produces a stable mixed equilibrium, with a fixed proportion of high- versus low-impact strategies and a fixed state of the environment. At this equilibrium, the average fitness attained by the population of forecasting types exceeds the average fitness that would arise in a population of myopic types. In other words, a population of pure forecasters can stabilize the eco-evolutionary system and thereby increase average public welfare, compared to myopic decision makers.

\subsubsection*{Coexistence of myopic and forecasting types}
\begin{figure}
    \centering
    \includegraphics[width = \textwidth]{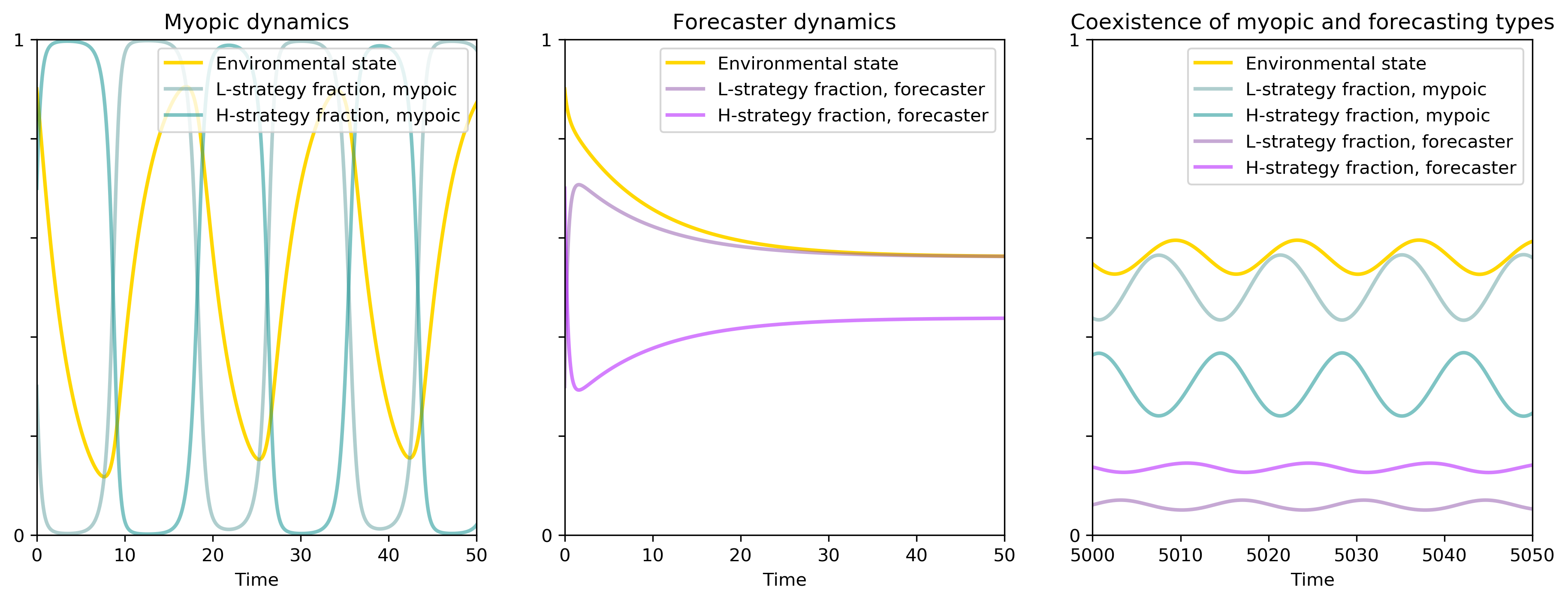}
    \put(-398,-10){(a)}
    \put(-242,-10){(b)}
    \put(-83,-10){(c)}
    \caption{(a) Eco-evolutionary game dynamics in a population of myopic types, who can each assume either the high-impact strategy ($H$) or the low-impact strategy ($L$). The feedback between strategic choices and the environment creates persistent cycles, in which the $L$ strategy is advantageous when the environmental state is high and the $H$ strategy is advantageous when environmental state is low. (b) If all individuals are forecasting types who integrate environmental forecasts into their strategy updating, then the state of the environment and the strategic frequencies reach a stable equilibrium. (c) Forecasting types can co-exist with myopic types. After a lengthy transient (see Figure~\ref{temporal}), both types co-exist in a stable limit cycle with reduced magnitudes of environmental and strategic variation compared to a population of myopic types alone. $\left(\epsilon_1=.3,~ \epsilon_2=.1,~ r=.15,~ C=5/1000,~ R_0=5,~ R_1=0,~ S_0=2,~ S_1=0,~ T_0=0,~ T_1=2,~ P_0=0,~ P_1=4\right)$}
    \label{temporal}
\end{figure}

When forecasting types and myopic compete and interact, the coupled system does not approach equilibrium as quickly. Figure~\ref{temporal} panel (c) shows the long-run dynamics of forecasting types and myopic types who can each assume either a high-impact strategy or a low-impact strategy. Since forecasting is costly, it is not possible for forecasters to displace the myopic agents altogether --  if they did, the environment would be completely stabilized, in which case the myopic agents make the same strategic choice as forecasters while avoiding the cost of forecasting. As a result, forecasting types and myopic types both persist, and so do environmental oscillations; but the amplitude of oscillations is diminished compared to what occurs in a population of myopic agents alone.

These results show that both cognitive types can coexist and that the presence of forecasting types, even when they comprise only a small fraction of the population, can  have a significant impact on the dynamics of the system and the fitness of both cognitive types. Despite the persistence of some environmental oscillations when forecasting and myopic types coexist, the resulting population mean fitness over one environmental cycle is much greater than when only myopic types are present, and it can even exceed the fitness when only forecasting types are present.

\subsubsection*{Invasion of forecasting types}
We have shown that forecasters can coexist with myopic types and have a large effect on strategic and environmental dynamics. But the question remains: can the forecasting type invade a population initially composed of myopic types and reach a substantial frequency? To study the invasion of forecasting, we initialize simulations with only myopic individuals and let this system relax to its steady-state, which features a stable limit cycle with large oscillations. We then perturb the system by introducing forecasters at a low frequency and simulate that invasion process.

For any particular set of parameters, we find that the likelihood of a successful invasion is greater when the initial frequency of forecasting types is greater. The success of invasion also depends on the strategy composition among the invading forecasters as well as the exact timing of the introduction of forecasters within the environmental cycle. 

Nonetheless, given sufficiently slow switching between forecaster and myopic types ($\epsilon_2<<1$), our simulations show that a successful invasion results in system-level behavior that always approaches the same limit-cycle, regardless of invasion timing and strategy mix. The long-run dynamics after forecasters successfully invade and coexist with myopic types are relatively simple: both types persist in a stable limit cycle with a relatively small amplitude of oscillation (Figure~\ref{temporal}c). But the invasion process itself is quite intricate. 

\begin{figure}
    \centering
    \includegraphics[width = \textwidth]{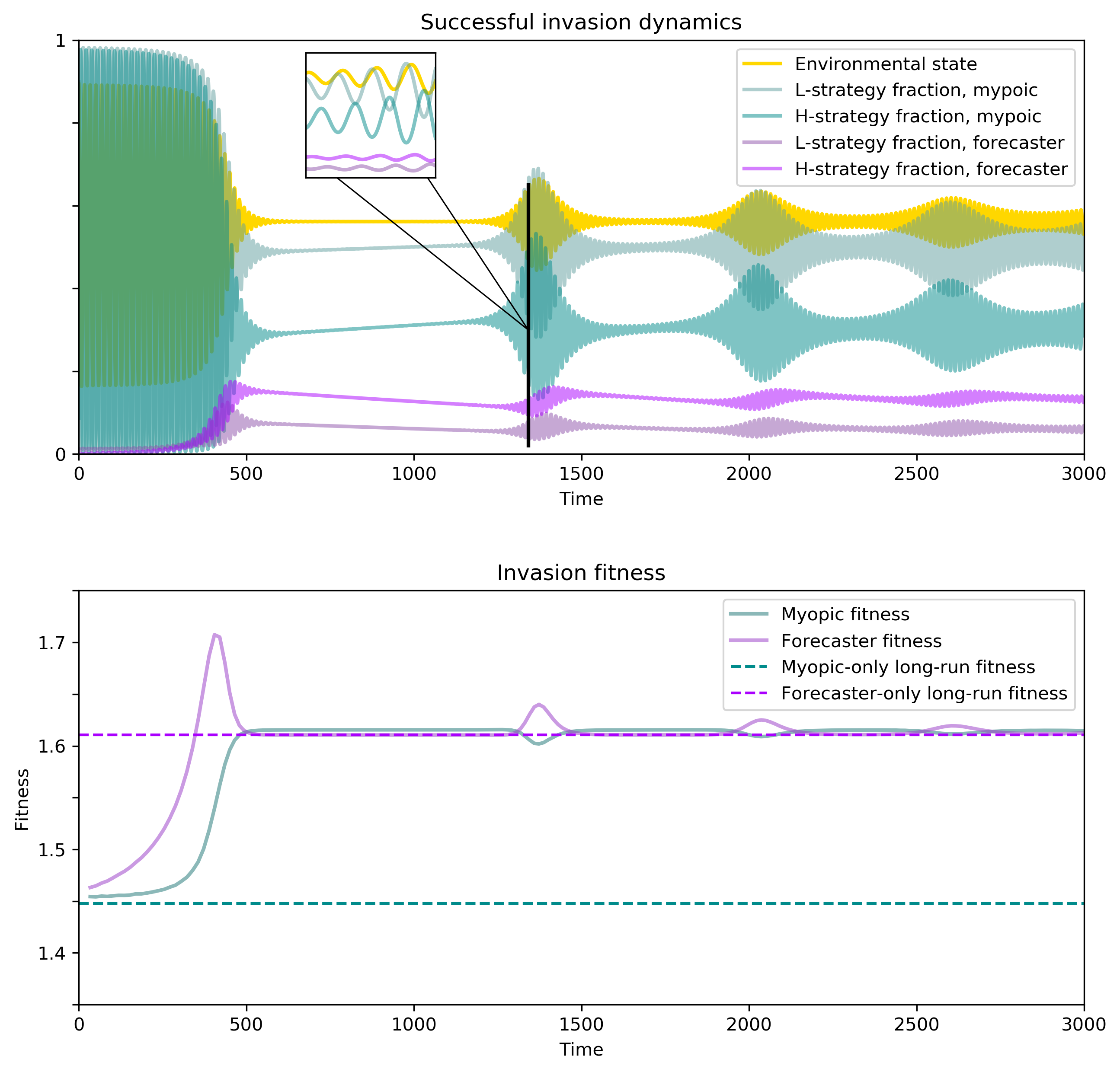}
    \put(-233,225){(a)}
    \put(-233,-10){(b)}
    \caption{(a) Environmental forecasting is assumed to be costly, but forecasting types can nonetheless invade a resident population of myopic types and eventually reach coexistence, while reducing the amplitude of oscillations.
    As this example shows, a successful invasion of forecasters takes orders of magnitude longer than a single environmental cycle. 
    (b) Dashed lines indicate the long-run fitness that would be attained by a population of purely myopic or purely forecasting types. Solid lines indicate the average fitness of forecasting and myopic sub-populations during the invasion process with both types present.
    The amplitude of environmental variability is reduced once forecasters reach appreciable frequency, which increases the fitness of forecasting and myopic types alike, eventually producing fitness as high as when only forecasters are present. Forecasting is thus a public good: it is individually costly and beneficial for all. 
    $\left(\epsilon_1=.3,~ \epsilon_2=.1,~ r=.15,~ C=5/1000,~ R_0=5,~ R_1=0,~ S_0=2,~ S_1=0,~ T_0=0,~ T_1=2,~ P_0=0,~ P_1=4\right)$ The initial frequency of forecasting types is $1/70$.}
    \label{invasionFitness}
\end{figure}

The dynamics of a successful invasion occur at a much slower timescale than the environmental oscillations (Figure~\ref{invasionFitness}a), so that the invasion process takes many environmental cycles before the forecasters reach an appreciable relative abundance. Figure~\ref{invasionFitness}b shows the dynamics of fitness for forecasting and myopic types (averaged over one environmental cycle). The dashed lines represent the long-run average fitness attained by a population composed entirely of myopic individuals or by a population of pure forecasters. For these parameters, when only forecasting types are present, the environment equilibrates to a fixed-point and this leads to a greater long-run fitness. When only myopic types are present, long-run average fitness is lower, and the persistent environmental oscillations are substantial. Forecasters can invade under highly variable environments because forecasting allows them to foresee and predict the best time to switch strategies ($L$ or $H$). This mitigates environmental oscillations and leads to a decrease in the magnitude of environmental variability. The reduction in environmental variability that forecasters provide acts as a rising tide that lifts all boats: when the environment is more stable, both forecasters and myopic individuals achieve payoffs as high as those that are attained when only forecasters are present. Since forecasting is individually costly and its environmental benefits are shared by all, forecasting is a public good.
Despite serving as a public good, we see that forecasting can nonetheless emerge via evolution and increase the fitness of both myopic and forecasting individuals, even when forecasters comprise only a small portion of the population over the long term ($\sim15\%$ in the example of Figure~\ref{phasePlane}).  

\subsubsection*{Successful and unsuccessful invasions}
Whether or not forecasters will successfully invade and reach a stable frequency depends on initial conditions. In the simulations described above, we assume that when forecasters arise, they have the same mix of strategies as the resident myopic population. Nonetheless, the timing of the invasion within the environmental cycle and the initial frequency of invading forecasters determine whether the invasion will ultimately be successful and lead to coexistence.  

\begin{figure}
    \centering
    \includegraphics[width = \textwidth]{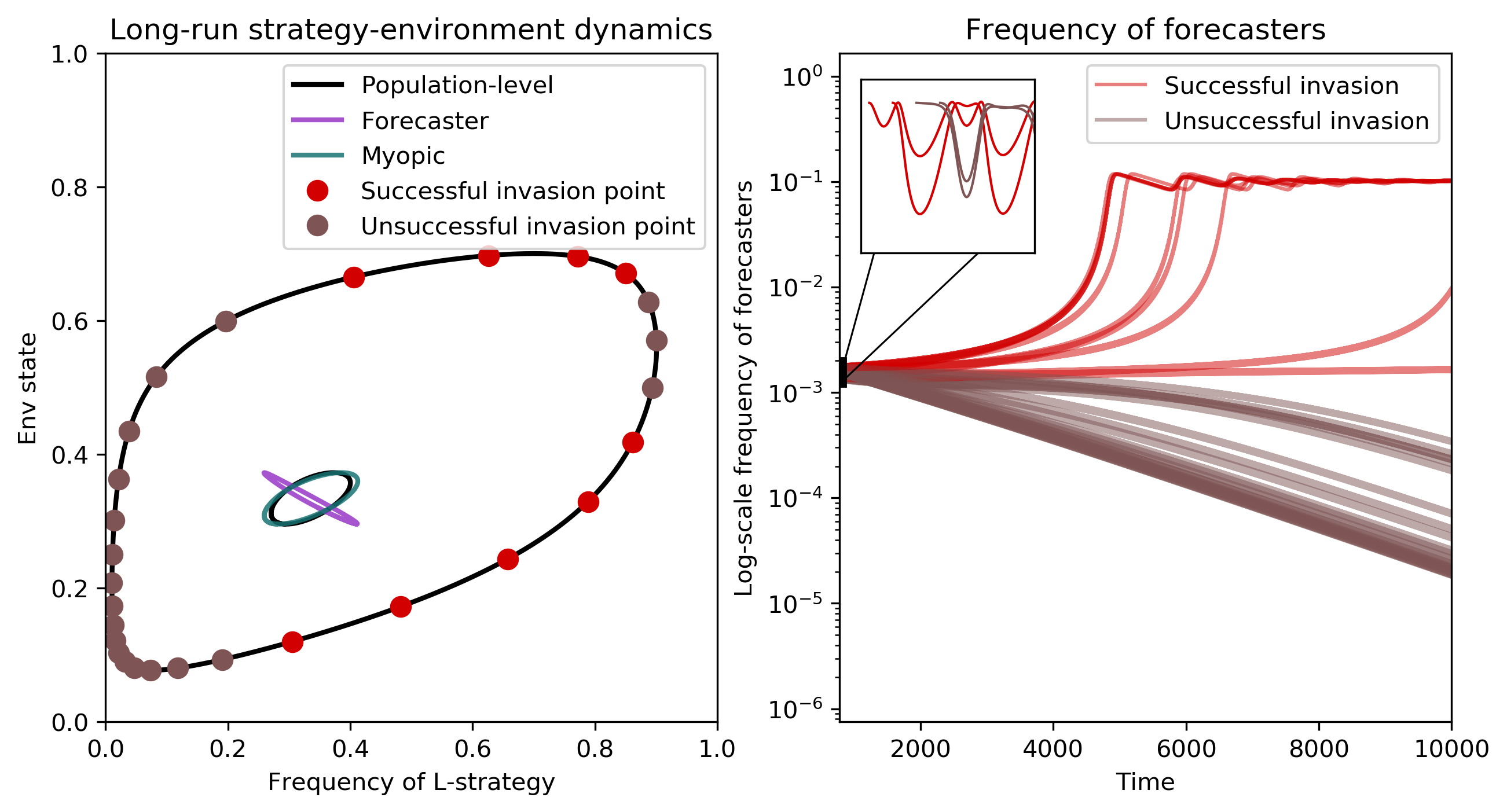}
    \put(-348,-10){(a)}
    \put(-120,-10){(b)}
    \caption{While forecasters can invade and coexist with  myopic types, a successful invasion is not assured and depends upon the timing of introduction. (a) In a purely myopic population,  eco-evolutionary game dynamics converge to the outer black orbit and proceed in a counter-clockwise direction. We study the potential invasion  of forecasting by introducing a small sub-population of forecasters at some point along this orbit. We assume that the forecasting sub-population has the same mix of strategies as the resident myopic population at the time of introduction. Points on the outer orbit indicate simulated introduction points that are evenly spaced in time. The color of the point illustrates whether or not the invasion of forecasters is successful from this point. The inner black orbit shows the long-run dynamics of the system, which are independent of the timing of the invasion provided it is successful. The purple and teal orbits illustrate the corresponding long-run dynamics among forecasting and myopic sub-populations. (b) Invasions progress very slowly, so that many environmental cycles pass before they approach the long-run dynamics illustrated in panel (a). The frequency of forecasters oscillates rapidly throughout this slow invasion process, and it may take a long time before it is clear whether the invasion will succeed or fail.$\left(\epsilon_1=3/10,~ \epsilon_2=1/10,~ r=5/100,~ C=5/1000,~ T_1=R_1=2,~ P_1-S_1=4,~ S_0-P_0=1,~ R_0-T_0=3\right)$ The initial frequency of forecasting types is $1/550$.}
    \label{phasePlane}
\end{figure}

Figure~\ref{phasePlane}a shows invasion points that are evenly spaced in time, within a phase plane plot. The outer black orbit represents the long-run dynamics of a population of purely myopic individuals,  with dynamics proceeding in a counter-clockwise direction. Whether or not a small frequency of forecasters can invade depends upon at what time point, within this periodic cycle, the forecasters are introduced: successful timings are shown as red dots and unsuccessful timings as brown dots. Even a successful invasion requires many environmental cycles to establish a stable frequency of forecasters (Figure \ref{phasePlane}b).

The inner black orbit in Figure \ref{phasePlane}a shows the long-run population-level dynamic after a successful invasion of forecasters. The magnitude of environmental and strategic variability is decreased substantially after invasion, which is reflected by a much smaller orbit in phase space.   
   
Whether or not forecasters can successfully invade also depends on how  much they value the present versus the future. In general, if forecasters place greater value on expected future payoffs (that is, a small discount rate $r$), this tends to increase their probability of successful invasion and establishment in a myopic resident population, all else equal (SI Figure~\ref{invasionSuccessRate}). Whereas if forecasters primarily value the present (large $r$), then they do not differ much in their strategic behavior compared to the myopic types, except they must pay a cost and so they cannot invade (SI Figure~\ref{invasionSuccessRate}). And so, placing great value on the future always benefits forecasters, even if they end up putting more decision-making weight on predictions for the far-future that may prove false. These results make intuitive sense considering that forecasters types are continually updating their forecasts based on the current trajectory of environmental change.

\section*{Discussion}
Strategic interactions often play out in changing environments.
But if the environment is always changing, will individuals who account for environmental dynamics be favored when making strategic decisions? We find that in many circumstances environmental forecasters, who pay a cost for their forecasting ability,  can indeed invade otherwise myopic populations engaged in eco-evolutionary games. Although a population of pure forecasters can produce a completely stable environment, forecasters cannot, in these cases, entirely over-take a population of myopic agents. Nonetheless, even when forecasters remain at low abundance, they have a striking effect on the dynamics of the eco-evolutionary game, greatly mitigating the amplitude of environmental variation and increasing fitness for all. 

Whereas classical bioeconomic approaches to decision-making assume that agents have complete information about the dynamics of the environment~\citep{clark}, evolutionary approaches often assume the opposite: individuals update their strategy based solely on present conditions~\citep{safarzynska2010}. Here, we analyze a middle ground and ask whether agents who act slightly more like those studied in bioeconomic approaches can emerge in a strictly myopic population. That this invasion can be successful means that evolution by natural selection, or myopic imitation, can promote the emergence of more sophisticated cognition. A surprising and important caveat is that forecasters cannot replace myopic individuals; instead, both types coexist. 

The invasion of forecasters in an eco-evolutionary game acts as a rising tide that lifts all boats, increasing everyone's fitness. In our model, forecasting behaves like a public good -- providing benefits to forecasters and myopic types alike -- and, thus, the invasion of forecasters can be thought of as a meta social dilemma, where both strategies ($L$ or $H$) and decision-making types (myopic or forecasting) exhibit features of a social dilemma. When a myopic population experiences cyclic dynamics, the long-run average payoffs are even lower than what would occur under the tragedy of the commons. This phenomenon has been referred to as an oscillating tragedy of the commons~\citep{weitz}, but given the damaging consequences for individuals' fitness, the `catastrophe of the commons' may be more apt. The invasion of forecasters resolves this catastrophe by ushering in greater environmental stability and increased fitness for forecasting and myopic types alike. 

Rationality is a bedrock assumption that underlies much of microeconomic theory. However, the extent to which economic rationality applies to human decision makers has long been questioned. Bounded rationality is an alternative model that acknowledges the limits of human cognition~\citep{simon1955}. Similarly, prospect theory accounts for decision-making heuristics that violate classical notions of rationality and alter behavior~\citep{tversky1974}. These advances were central to the formation of behavioral economics, which explores how economic, social, and cultural factors effect decision-making. More recent work has considered the potential for multiple modes of decision-making to coexist in a population or within a single individual~\citep{kahneman2011thinking,rand2017}. Models starting from an evolutionary perspective have integrated foresight, where agents consider how changes in their behavior may alter the choices of others in the future~\citep{perry2020foresight,perry2018collective}. However, little work has explicitly analyzed the consequences of feedback among modes of decision-making, dynamics of behavior, and the environmental setting, with a few notable exceptions~\citep{adamson2020resource,Austrup}.     

The focus of our paper has been on human-environmental systems, yet our modeling approach may possibly be applied to non-human organisms that live in variable environments. Flowering plants are most successful when they time their blossoming to align with favorable environmental conditions, such as temperature, pollinator abundance, and density of competitors. Accounting for environmental cues, including environmental change, may help plants achieve optimal flowering timing \citep{vermeulen2015}. Migratory species face a similar dilemma about when to leave one habitat and head for another \citep{johansson2012}; and some form of environmental forecasting my be advantageous in this context. Female Dusky Warblers modify their nest site choice in response to changes in the density of predators~\citep{forstmeier}. This could generate feedback between the birds' nesting strategies and the predation environment that the birds face. Likewise, Tengmalm's owls in Western Finland seem to adjust their clutch sizes in accordance to the 3- year population cycle of voles, their prey, in the area  \citep{korpimaki1991}. In each of these settings, our model suggests that individuals with the ability to forecast environmental change, even if this capability comes at a cost, may be favored by selection.

While we have studied the evolution of a new mode of decision-making that forms  beliefs about future environmental states, a closely related phenomenon called theory of mind falls outside the scope of our analysis. Theory of mind describes the ability to conceptualize the way in which others make decisions~\citep{apperly2012}. Future work could study forecasting of others' strategy dynamics, in addition to or in lieu of environmental forecasting. We expect that strategic forecasting may have qualitatively distinct population-level impacts than environmental forecasting, and it may be favored in different settings. Additionally, here we have analyzed a well-mixed model where all individuals of all types and strategies interact; in some contexts, a hierarchical model might be more realistic. In Supplementary Information Section~\ref{altDynam}, we describe a hierarchical model where forecasting types and myopic types form two distinct sub-populations that compete with each other at the population level. The hierarchical model produces dynamics of greater complexity than the well-mixed model presented in the main text.

Rather than predicting a steady advance in the sophistication of cognition, our analysis suggests that a more likely outcome is the coexistence of multiple modes of decision-making. This finding aligns with results from psychology that indicate  most people have automatic modes of decision-making as well as more deliberative, and mentally taxing, modes \citep{kahneman2011thinking,rand2017}. Our finding of coexistence between forecasting and myopic types provides theoretical support for behavioral-economic hypotheses about how decisions are actually made. Our results help explain why automatic decision-making persists, and they describe a context in which higher-level cognition can be favored evolutionarily while also bringing public benefits to all cognitive types.

\section*{Acknowledgements}
The findings and conclusions in this publication are those of the authors and should not be construed to represent any official USDA or U.S. Government determination or policy. Code are publicly available at \url{https://github.com/atilman/EvolutionForecasting}

\bibliographystyle{apalike}
\bibliography{bibli}

\renewcommand{\figurename}{SI Figure}

\setcounter{figure}{0}
\appendix
\newpage

\begin{center}
\section*{The evolution of forecasting \\
for decision-making in dynamic environments\\ 
\textit{Supplementary Information}}
\end{center}

In the main text, we describe and analyze an eco-evolutionary model to explore the emergence of forecasting types in dynamically variable environments. The model describes competition among forecasting and myopic types based on mass action kinetics. In other words, the populations of all four states of agents (high- or low-impact strategies and myopic or forecasting types) are well mixed, and interactions occur proportional to frequency. An alternative formulation of the model could consider a hierarchical structure, where forecasting and myopic sub-populations update their strategies within their type, and the sub-populations of cognitive types compete at the group level. 

Here, we present the alternative, hierarchical formulation of the model, and we illustrate some of the complex dynamics that this alternative formulation can generate. We perform a change of variables so that the hierarchical and mass-action formulations of the model can be directly compared. This comparison reveals that the primary difference between the models is how the speed of intra-type dynamics responds to the frequency of a type.

Next, we present additional findings for the mass-action model. We illustrate the importance of the discount rate for invasion success, and find that the more weight the future has in decision-making, the more likely forecasters are to invade. We show that forecasting does not entirely resolve the tragedy of the commons: optimal population-level fitness is not necessarily achieved under dominance by forecasting types.

Finally, we introduce an individual-based model that is converges to the model we consider in the main text in the limit of large population size and weak selection. 

\medskip

\section{Hierarchical model of forecaster emergence}\label{altDynam}
In this section, we consider an alternative eco-evolutionary dynamic, where there is a hierarchical structure to the system. Forecasters interact only with forecasters, and myopic individuals interact only with myopic individuals for determining strategy dynamics. Then, forecasting and myopic populations compete, based on present payoffs. This hierarchical dynamic leads to a different system of equations governing system dynamics, resulting in some qualitatively distinct outcomes. 

We can write the dynamics of this hierarchical system as

\begin{align}
\dot{y}_L^m&=y_L^m(1-y_L^m)]\left(\pi_L-\pi_H\right)\\
\dot{y}_L^f&=y_L^f(1-y_L^f)\left(f_L-f_H\right)\\
\dot{z}^f&=\epsilon_2 z^f(1-z^f)\left[(y_L^f-y_L^m)\left(\pi_L-\pi_H\right)-C\right]\\
\dot{n}&=\epsilon_1\left[y_L^f z^f+y_L^m(1-z^f)-n\right].
\end{align}
Notice that $\dot{n}$ also appears in the dynamical equation for the frequency of strategy $L$ among forecasters, this is because forecasters use the current rate of change of the environment to forecast likely future environmental states. Further, notice that $C$ is the cost of forecasting. We assume that both myopic and forecasting agents update their strategies on the same timescale, and that $\epsilon_2$ and $\epsilon_1$ are the relative timescales (as compared to strategy dynamics) of switching between forecasting and myopic, and environmental dynamics, respectively. In sum, we have a four-dimensional system with each variable bound between 0 and 1.
\paragraph{Equation for $\dot{y}^m_L$ }
Myopic agents update their strategy according to standard replicator dynamics. The payoff difference between the strategies controls the dynamics of switching.

\paragraph{Equation for $\dot{y}^f_L$}
Forecasting agents update their strategy based on the present state of the systems as well as based on how the changes in the environment will change future payoffs. Taking into account this future modifies the equation for selection in this case, with forecasted payoffs utilized for strategy dynamics.

\paragraph{Equation for $\dot{z}^f$} While we assumed that forecasters update their their strategy according to predicted future outcomes, however, forecaster must nonetheless compete with myopic agents in the present. Thus the frequency of forecasters evolves in response to the present payoffs of myopic and forecasting types. The payoff to a myopic agent is 

\begin{equation}
    y_L^m\pi_L+(1-y_L^m)\pi_H,
\end{equation}
and the payoff to forecasting agents is

\begin{equation}
    y_L^f\pi_L+(1-y_L^f)\pi_H-C,
\end{equation}
Where $\pi_L$ and $\pi_H$ are the payoffs of strategies 1 and 2. With these, we can write the payoff difference between forecasting agents and myopic agents as

\begin{equation}
      (y_L^f-y_L^m)\pi_L+(1-y_L^f-1+y_L^m)\pi_H-C=(y_L^f-y_L^m)(\pi_L-\pi_H)-C.
\end{equation}
Intuitively, when strategy $L$ yields higher payoffs than strategy $H$ given the state of the system, forecasters will have higher payoffs than myopic agents if the frequency of strategy $L$ among forecasters, $y_L^f$, is greater than among myopic agents, $y_L^m$. Further, since forecasting is costly, the benefits of forecasting must outweigh the cost, $C$, for forecasters to increase in frequency. 

\paragraph{Equation for $\dot{n}$} This is a straightforward modification of the decaying resource environmental feedback from Tilman et al. (2020) with the frequency of strategy $L$ written in terms of our variables. 

\medskip
Comparing the equations for the pairwise interaction (mass action) model with the equations that arise from the hierarchical model provides insight into how the two model structures diverge, and why. In both models, the dynamical equations that govern the frequency of forecasters, $z^f$, are identical. In both models forecasting and myopic types compete based on present fitness. This does not imply that the invasion processes withing the two models are equivalent, however. 

In the pairwise interaction model, when forecasters are rare, they mostly interact with myopic individuals, this limits the utility of forecasting, since the forecast is only utilized when interacting with another forecaster. This accounts for the existence of the $z^f(f_L-f_H)$ term in forecaster strategy dynamics under pairwise interaction; when forecasters are rare, $z^f<<1$, $f_L-f_H$ has a minimal impact on the strategy dynamics of forecasters. 

Under the hierarchical model, in contrast, there is no effect of $z^f$ on strategy dynamics for forecasting or myopic types. This is because there are assumed to be two well mixed sub populations where the rate of interaction (and thus strategy dynamics) does not depend on $z^f$.  

Qualitatively, the two models produce distinct invasion dynamics. As shown, invasion under the pairwise interaction model occurs with a long transient, where the frequency of forecasting stays very low. In the long run, a successful invasion by forecasters leads to a new limit cycle. 

\begin{figure}
\centering
\hspace{0mm}\vspace{0mm}
\subfloat[Invasion at $t=105$]{
         \begin{overpic}[width=0.45\textwidth]{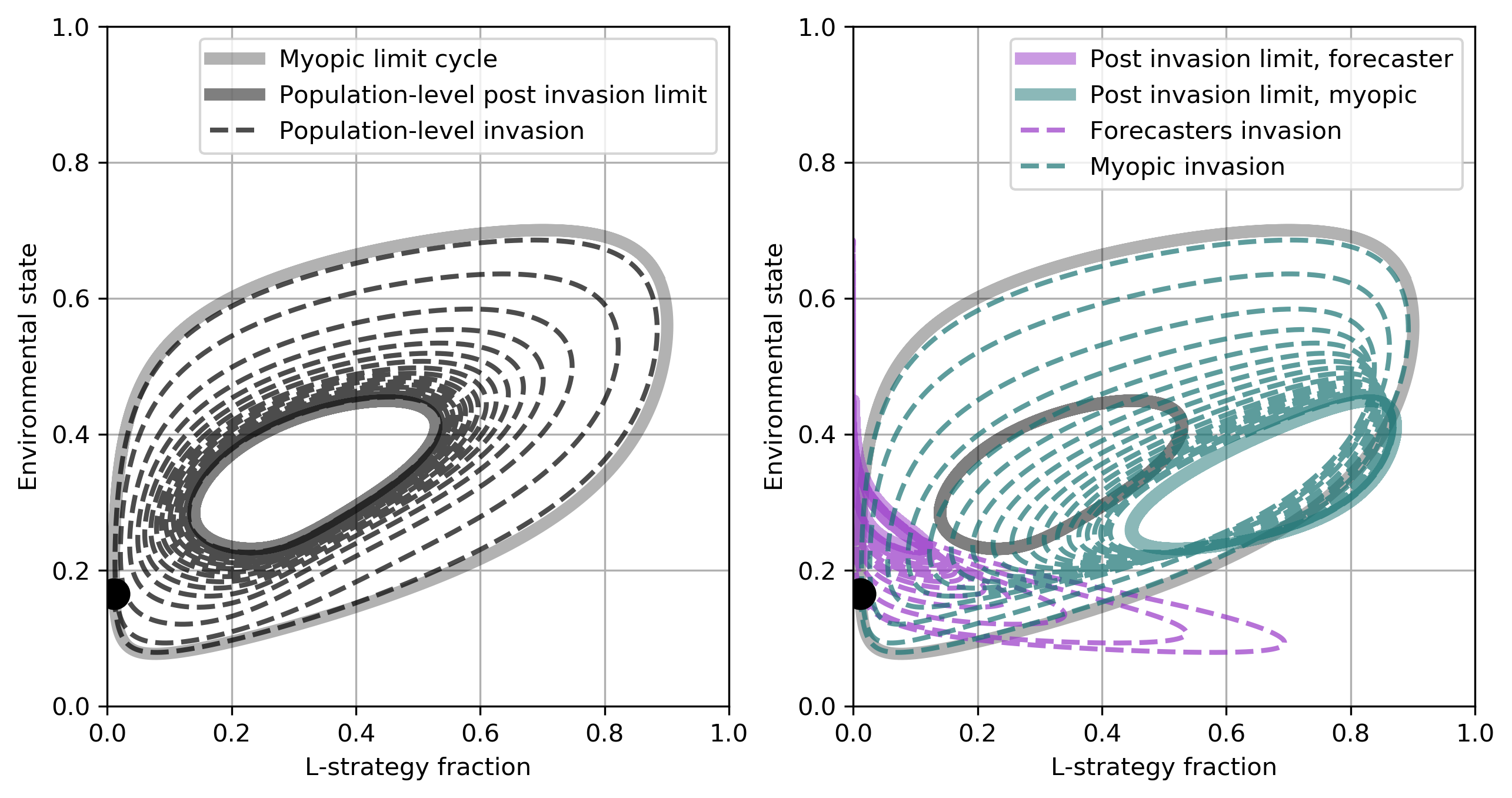}
       \end{overpic}
}\\
\subfloat[Invasion at $t=100$]{
  \begin{overpic}[width=0.45\textwidth]{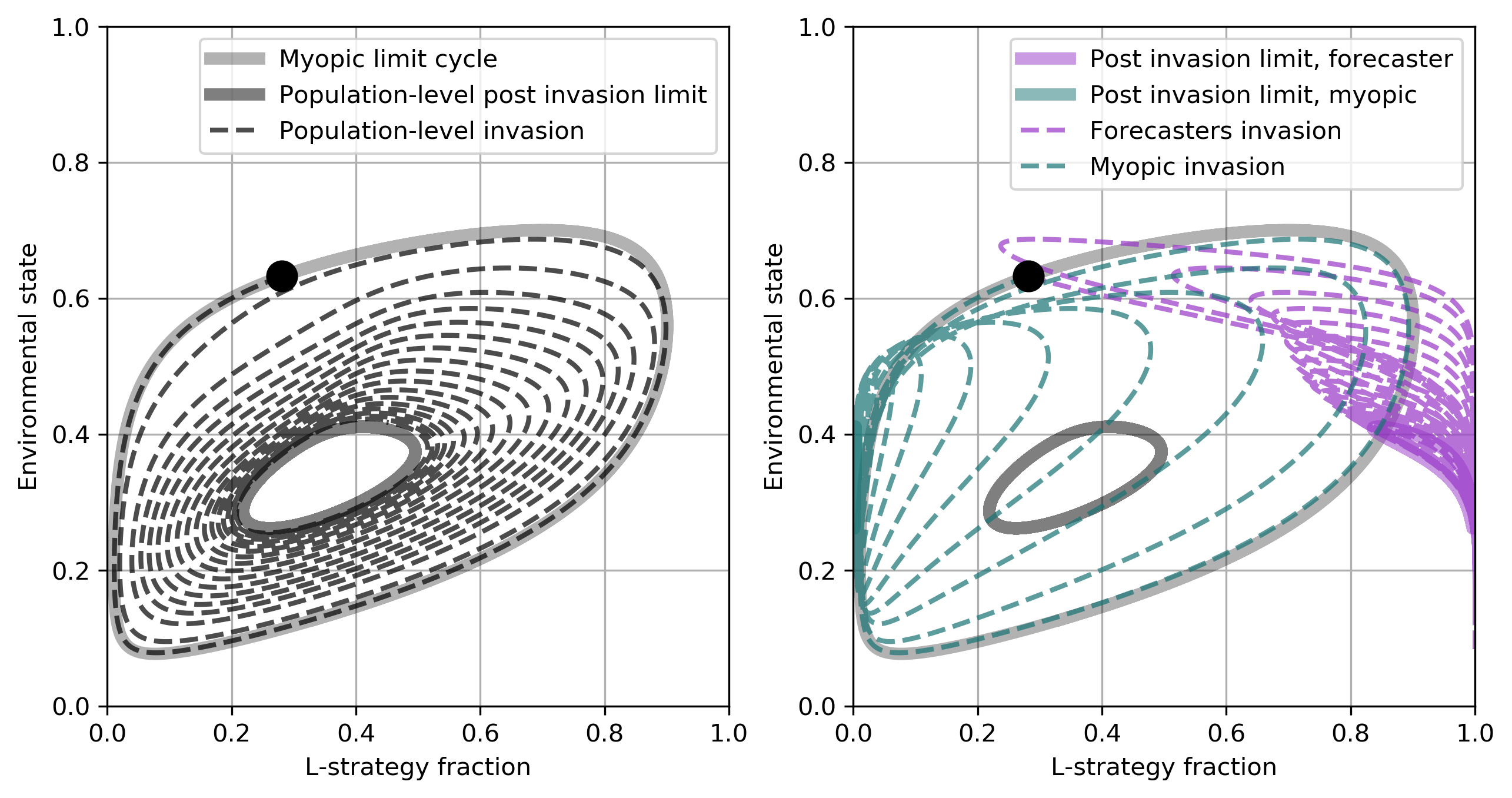}
       \end{overpic}
}\\
\subfloat[Invasion at $t=95$]{
  \begin{overpic}[width=0.45\textwidth]{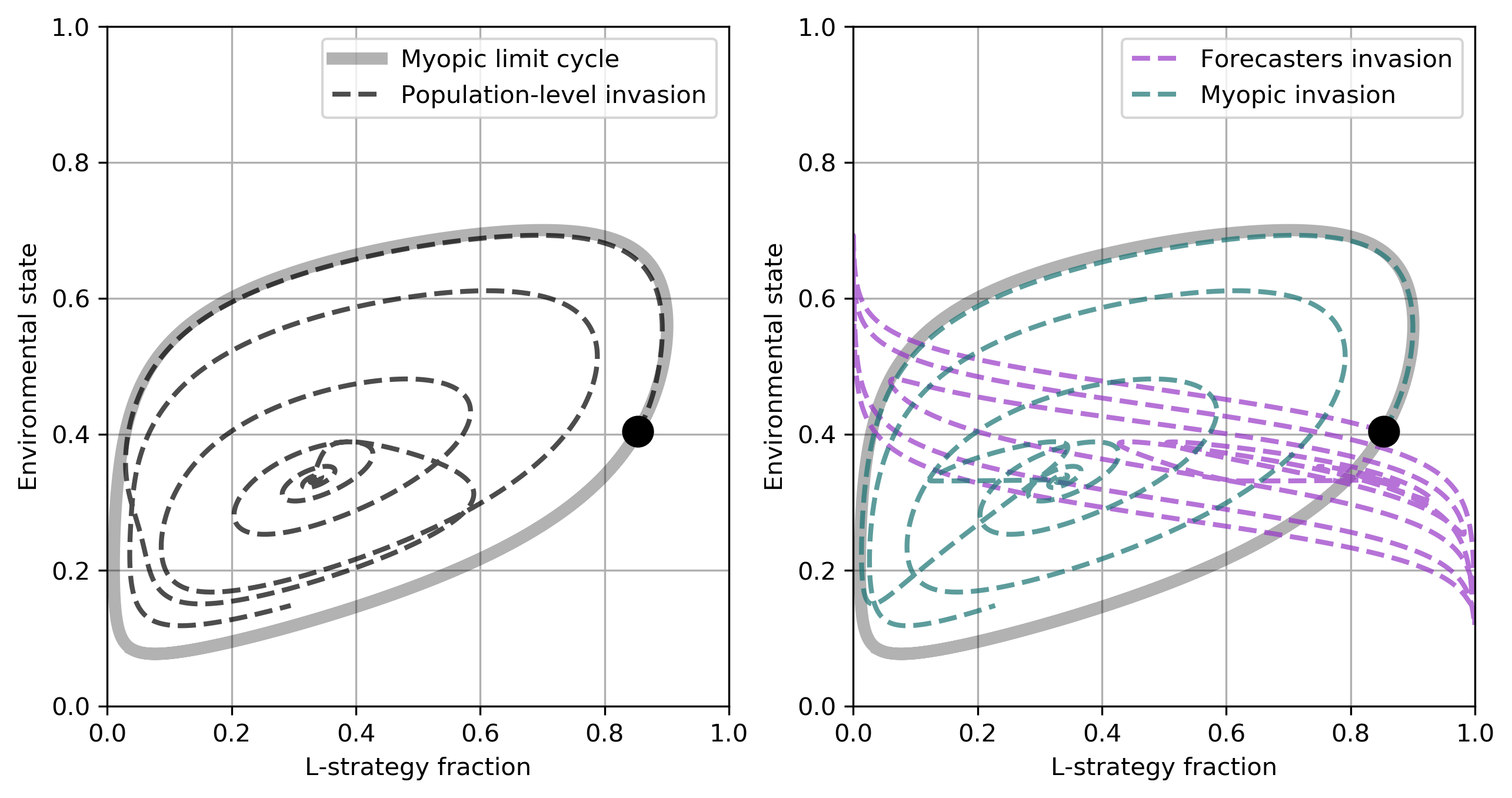}
       \end{overpic}
}
\caption{The three panels show that the hierarchical model produces different long-run dynamics given different system states when forecasting types are introduced. In these examples, forecasting types match the frequency of strategy $L$ and $H$ among the myopic types at the time of invasion. The outer light grey loop in the panels is the limit cycle reached by a population of myopic individuals. The dashed lines represent population-level and within type (forecasting or myopic) strategy-environment dynamics during the invasion process. Light red and light blue loops are the long run dynamics of myopic and forecasting types, respectively. Notice that the post-invasion long run dynamics vary depending on the state of the system at the time of invasion. Panels (a) and (b) show subtly different long-orbits post invasion. Panel (c) shows a case where long-run dynamics do not settle into a simple periodic orbit, but rather exhibit periods of environmental (and strategy) stability followed by periods of environmental (and strategy) variability.  
$\left(\epsilon_1=3/10,~ \epsilon_2=1,~ r=15/100,~ C=1/100,~ T_1-R_1=2,~ P_1-S_1=4,~ S_0-P_0=1,~ R_0-T_0=3\right)$ The initial frequency of forecasting types is $1/100$.}

\label{inv_timing}

\end{figure}

Under the hierarchical model, forecasters can increase in frequency much more rapidly, since their strategy dynamics are not slowed when they are rare. SI Figure~\ref{inv_timing} shows that the hierarchical model produces more complex long run dynamics. For a single set of parameters, a range of long run behaviors are possible. It appears as though there may exist a stable manifold upon which a continuum of neutrally stable orbits occur. This finding, however, is without proof. 

\section{Relationship between hierarchical and mass-action models}
We have described the hierarchical model with a different coordinate system than we used to describe the mass action model in the main text. Here, we perform a change of coordinates on the mass-action model so that we can directly compare the two models, and consider some possible sources of divergence between their results. First, recall the system of equations analyzed in the main text, given by

\begin{align*}
\dot{z}_L^m &= z_L^m z_H^m (\pi_L-\pi_H) + \epsilon_2 z_L^m z_L^f C + \epsilon_2 z_L^m z_H^f (\pi_L-\pi_H+C)\\
\dot{z}_H^m &= - z_H^m z_L^m (\pi_L-\pi_H) + \epsilon_2 z_H^m z_L^f (-\pi_L+\pi_H+C) + \epsilon_2 z_H^m z_H^f C\\
\dot{z}_L^f &= z_L^f z_H^f (f_L-f_H) - \epsilon_2 z_L^f z_L^m C + \epsilon_2 z_L^f z_H^m (\pi_L-\pi_H-C)\\
\dot{z}_H^f &= -z_H^f z_L^f (f_L-f_H) + \epsilon_2 z_H^f z_L^m (-\pi_L+\pi_H-C) - \epsilon_2 z_H^f z_H^m C\\
\dot{n}_{~} &= \epsilon_1(z_L^f+z_L^m-n).
\end{align*}
where $\dot{z}_L^m+\dot{z}_H^m+\dot{z}_L^f+\dot{z}_H^f=1$, denoting the population-wide frequencies of the four strategic types, and $n\in[0,1]$ is the state of the environment.

With the following transformations, we can perform a change of coordinates to express the mass-action model in the same terms as the hierarchical model.

\begin{align}
    y_L^m &= \frac{z_L^m}{z_L^m+z_H^m}\\
    y_L^f &= \frac{z_L^f}{z_L^f+z_H^f}\\
    z^f &= z_L^f+z_H^f\\
    n &= n.
\end{align}
With this change of coordinates we can write the dynamical system as

\begin{align}
    \dot{y}_L^m &= \frac{\partial y_L^m}{\partial z_L^m}\dot{z}_L^m + \frac{\partial y_L^m}{\partial z_H^m}\dot{z}_H^m\\
    \dot{y}_L^f &= \frac{\partial y_L^f}{\partial z_L^f}\dot{z}_L^f + \frac{\partial y_L^f}{\partial z_H^f}\dot{z}_H^f\\
    \dot{z}^f &= \dot{z}_L^f+\dot{z}_H^f\\
    \dot{n} &= \dot{n}.
\end{align}
Through collection of terms and substitution of variables, we can write the system in our new coordinate system as 

\begin{align}
    \dot{y}_L^m &= y_L^m\left(1-y_L^m\right)\left[(1-z^f)(\pi_L-\pi_H)+\epsilon_2 z^f (\pi_L-\pi_H)\right]\\
    \dot{y}_L^f &= y_L^f\left(1-y_L^f\right)\left[z^f(f_L-f_H)+\epsilon_2(1-z^f)(\pi_L-\pi_H)\right]\\
    \dot{z}^f &= \epsilon_2 z^f (1-z^f) \left[(y_L^f-y_L^m)(\pi_L-\pi_H)-C\right]\\
    \dot{n}_{~} &= \epsilon_1 \left[y_L^f z^f+y_L^m(1-z^f)-n\right].
\end{align}
This coordinate systems highlights a few important properties about the system, as constructed. First, the $z^f$ equation shows how forecasting can be favored at the population level. Forecasting types increase in frequency when the forecasting sub-population has a higher frequency of the favored strategy. This is achieved when forecasters foresee that the optimal strategy will soon change, and adopt that strategy earlier and more rapidly than myopic types. Nonetheless, myopic individuals always are at an advantage, as they avoid paying the cost $C$, of having the ability to forecast. 

Second, notice that the forecasting terms, $f_L$ and $f_H$, only appear in the forecaster strategy equations, and when the prevalence of forecasting is low ($z_L<<1$), little weight is given to these forecasts. Instead, forecasters updating is dominated by the switching of myopic types to forecasters, but this switching is slow and cannot overcome the cost of forecasting, $C$. Simulations indicate that forecasters need a critical mass to invade, indicating bi-stability. For $\epsilon_2=1$, inspection of the dynamical equations near $z^f=0$ seems to confirm this, forecasting types have the same dynamics as myopic types, but pay a fixed cost $C$. While forecasting has minimal effect when the frequency of forecasting types is low, we nonetheless have cases where forecasting types can invade. 

Third, the strategy dynamic of myopic types is impacted relatively less by changes in the abundance of myopic individuals ($1-z^f$). Within- and cross-type switching is governed by the same process, compensating somewhat for $\epsilon_2$'s slowing of strategy dynamics.

\section{Additional mass-action model results}
We describe the oscillatory dynamics that occurs under a population composed entirely of myopic types as a `catastrophe of the commons'. This is because the oscillations that occur lead to further declines in average fitness or profit for the population that would be expected given a standard tragedy of the commons. Forecasting can mitigate or eliminate these oscillations, but when forecasting is costly, environmental stability cannot be attained when forecasting and myopic types compete. Nonetheless, the fitness gains of forecasting are nevertheless largely realized. 

\begin{figure}[htpb]
    \centering
    \includegraphics[width = \textwidth]{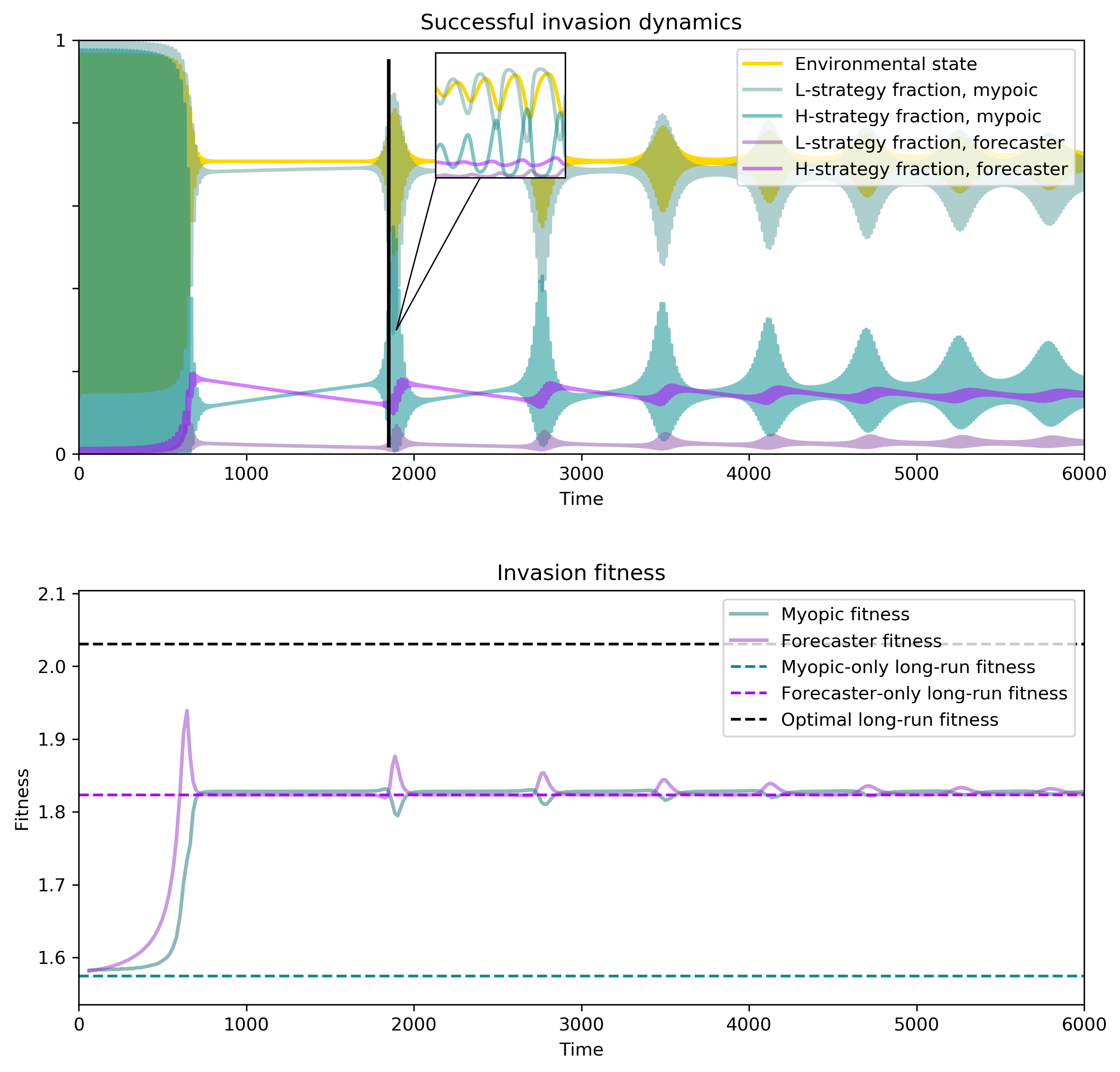}
    \put(-233,225){(a)}
    \put(-233,-10){(b)}
    \caption{(a) Forecasting types can invade a resident population of myopic types, leading to coexistence while reducing the amplitude of oscillations.
    As this example show, a successful invasion of forecasters takes orders of magnitude longer than a single environmental cycle. 
    (b) Dashed lines indicate the long-run fitness that would be attained by populations of purely myopic types, purely forecasting types, or the maximum level of sustained fitness that could be attained at the population level. Solid lines indicate the average fitness of forecasting and myopic sub-populations during the invasion process with both types present.
    The dynamics of forecasting and myopic types is a social dilemma, forecasting types create public benefits for both forecasting and myopic types. Remarkably, the possible fitness gains of forecasting are attained even in the face of this dilemma.
    Forecasting mitigates but does not resolve the tragedy of the commons. The optimal long-run population average fitness is greater than the fitness that a population of forecasting types enjoy. $\epsilon_1=0.3,~ \epsilon_2=0.1,~ r=0.15,~ C=0.005,~ R_0=8,~ R_1=0,~ S_0=2,~ S_1=0,~ T_0=0,~ T_1=2,~ P_0=0,~ P_1=4$. The initial frequency of forecasting types is $1/70$. 
    }
    \label{optimalFitness}
\end{figure}

However, this does not imply that forecasting types entirely resolve the tragedy of the commons. SI Figure~\ref{optimalFitness} shows that the optimal fitness that could be attained by a population -- by exogenously fixing the frequency of the two strategic types -- exceeds the population fitness under forecasting types alone, or under the coexistence of forecasting and myopic types. Therefore, forecasting can resolve a `catastrophe of the commons' but not the tragedy of the commons. Additional mechanisms are required to resolve problems of cooperation.

\begin{figure}[htpb]
    \centering
    \includegraphics[width=.9\textwidth]{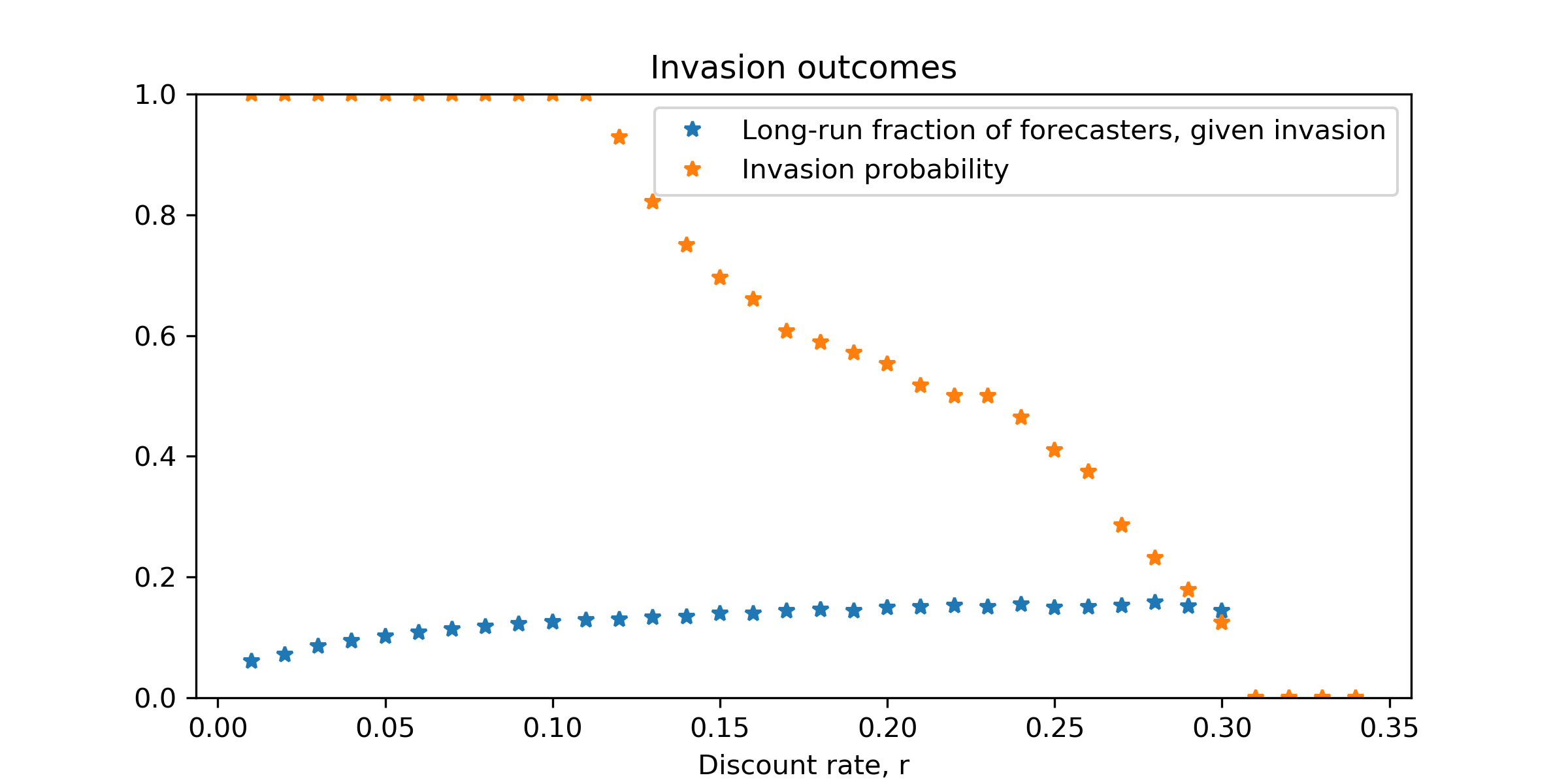}
    \caption{All else being equal, the lower the discount rate, $r$, the more likely it is that forecasting can invade. This implies that valuing the future more always benefits forecasters, even if they end up putting more decision-making weight on predictions for the far-future that may prove false. However, there is a non-monotonic relationship between the long-run frequency of forecasters and the discount rate that they employ.  $\left(
\epsilon_1 = 3/10, 
\epsilon_2 = 1/10, 
C = 5/1000, 
T_1-R_1 = 2, 
S_0-P_0 = 1, 
R_0-T_0 = 3, 
P_1-S_1 = 4\right)
$ Invasion strategy matched resident. The initial frequency of forecasting types is $1/100$.}
    \label{invasionSuccessRate}
\end{figure}

A key parameter associated with forecasting types is their discount rate. The discount rate determines how much weight forecasters place on the future in decision-making, and so it alters their effective time horizon. SI Figure~\ref{invasionSuccessRate} shows that as the discount rate decreases, and forecasting types care more about the future, their invasion success rate increases. Lower discount rates are also associated with a small decrease in the long-run frequency of forecasting types after invasion. In contrast to Adamson and Hilker (2020), we do not find that caring too much about the distant future can backfire and lead to the reemergence of oscillatory dynamics. This is likely because forecasting types update their projections and assessments continuously. Incorrect predictions about the distant future need to not be adhered to by forecasting types. As environmental trends reverse, so do forecasters predictions.

\section{Individual-based model}\label{IBM}

In this section, we present an individual-based model that converges to the model we consider in the main text as a limiting case. Consider a well-mixed population of $Z$ individuals interacting in an environmental state, $n$. Within a timestep $\tau$ an agent randomly interacts with another individual in the population and decides whether to adopt that other individual's strategy and (potentially) their decision-making type. In the same timestep, the environment responds to the strategy mixture in the population.

Let $Z^m_L$, $Z^m_H$, $Z^f_L$, and $Z^f_H$ be the number of myopic low-impact strategists, myopic high-impact, forecaster low-impact, and forecaster high-impact strategy individuals in the population, respectively. 

\subsection{Intra-type strategy change}
Myopic individuals will solely consider changing their strategy when they meet other myopic types with a different strategy. They compare the instantaneous payoff of their strategy, $\pi_\text{current}$, with that of the alternative strategy, $\pi_\text{alterante}$, and change with a probability given by a sigmoid function,

\begin{equation}
    p(x)=\frac{1}{1+\exp{\left[-2\beta x\right]}},
\end{equation}

as $p(\Pi_\text{alternate}-\Pi_\text{current})$. Notice that $p(0)=1/2$, and that this function has the special property that 
\begin{equation}
p(x)-p(-x) = \tanh(\beta x)
\end{equation}
Also, note that $\beta$ controls the intensity of selection of the best strategy. Alternatively,  one can also think of $1/\beta$ as the degree of uncertainty on the payoff difference.

We can write, for a given time $t$, the probability a myopic individual changes from the low to high impact strategy within a timestep $\tau$, corresponding to a transition, at the population level, from the state $\{Z^m_L, Z^m_H, Z^f_L, Z^f_H, n\}$ at time $t$ to $\{Z^m_L-1, Z^m_H+1, Z^f_L, Z^f_H, n'\}$ at time $t+\tau$, as

\begin{equation}
        \tau T^{mL\downarrow mH\uparrow} =
        \tau\frac{Z^m_L}{Z} \frac{Z^m_H}{Z-1} p(\pi_H-\pi_L).
\end{equation}

This can be read as the probability that the individual considering the strategy change is a myopic low-impact one, that they choose an myopic high-impact to compare themselves to, and they do it with the probability $p$ discussed. For simplicity, there is a term that we omit that corresponds to the change in the environment which should read $\delta\left(n'-n(t)-\int^{t+\tau}_{t}\dot{n}(t')dt'\right)$, where $\delta(x)$ is a Dirac function, which simply states that the environment will evolve to a precise new value in that timestep. 

Identically, the probability a myopic individual changes from high to low extraction strategy within timestep $\tau$, corresponding to a transition, at the population level, of the state $\{Z^m_L, Z^m_H, Z^f_L, Z^f_H, n\}$ to $\{Z^m_L+1, Z^m_H-1, Z^f_L, Z^f_H, n'\}$, is

\begin{equation}
        \tau T^{mH\downarrow mL\uparrow} =
        \tau\frac{Z^m_L}{Z} \frac{Z^m_H}{Z-1}  p(\pi_L-\pi_H).
\end{equation}

The probability a forecaster changes from high to low or low to high extraction strategy within timestep $\tau$, will depend on discounted forecasted payoffs, $f$. This strategy change  corresponds to a transition of the state $\{Z^m_L, Z^m_H, Z^f_L, Z^f_H, n\}$ to $\{Z^m_L, Z^m_H, Z^f_L-1, Z^f_H+1, n'\}$ or $\{Z^m_L, Z^m_H, Z^f_L+1, Z^f_H-1, n'\}$. The probabilities of these transitions are given by

\begin{align}
\tau T^{fL\downarrow fH\uparrow} &=
        \tau\frac{Z^f_L}{Z}\frac{Z^f_H}{Z-1} p(f_H-f_L), \\
\tau T^{fH\downarrow fL\uparrow} &=
        \tau\frac{Z^f_L}{Z}\frac{Z^f_H}{Z-1} p(f_L-f_H).
\end{align}

\subsection{Inter-type change}

Whereas the dynamics within-types of individuals only considers the possibility of a change of strategy, the dynamics between individuals of different types may or may not also entail strategy change.   
When a myopic individual interacts with a forecasting type, they may switch type (and strategy, if necessary) according to a sigmoid function 

\begin{equation}
    \hat p(x)=\frac{1}{1+\exp{\left[-2\gamma x\right]}},
\end{equation}

so that the strength of selection within types and across types can be varied independently.

We can write the probability a myopic individual with a low impact strategy is replaced by a forecasting individual with a low impact strategy within timestep $\tau$, corresponding to a transition of the state $\{Z^m_L, Z^m_H, Z^f_L, Z^f_H, n\}$ to $\{Z^m_L-1, Z^m_H, Z^f_L+1, Z^f_H, n'\}$, as

\begin{equation}
        \tau T^{mL\downarrow fL\uparrow} =
        \tau\frac{Z^m_L Z^f_L}{Z(Z-1)} \hat p(-C),
\end{equation}
because the only payoff difference between these individuals is the extra cost, $C$, that the forecasting type incurs.
Identically for all other replacements

\begin{align}
\tau T^{mL\downarrow fH\uparrow} &=
        \tau\frac{Z^m_L Z^f_H}{Z(Z-1)} \hat p(\pi_H-\pi_L-C),\\ 
\tau T^{mH\downarrow fL\uparrow} &=
        \tau\frac{Z^m_H Z^f_L}{Z(Z-1)} \hat p(\pi_L-\pi_H-C),\\
\tau T^{mH\downarrow fH\uparrow} &=
        \tau\frac{Z^m_H Z^f_H}{Z(Z-1)} \hat p(-C),\\
\tau T^{fL\downarrow mL\uparrow} &=
        \tau\frac{Z^f_L Z^m_L}{Z(Z-1)} \hat p(C),\\
\tau T^{fL\downarrow mH\uparrow} &=
        \tau\frac{Z^f_L Z^m_H}{Z(Z-1)} \hat p(\pi_H-\pi_L+C),\\
\tau T^{fH\downarrow mL\uparrow} &=
        \tau\frac{Z^f_H Z^m_L}{Z(Z-1)} \hat p(\pi_L-\pi_H+C), \text{and}\\
\tau T^{fH\downarrow mH\uparrow} &=
        \tau\frac{Z^f_H Z^m_H}{Z(Z-1)} \hat p(C).\\
\end{align}

\subsection{Population state transitions}
The state of the system is fully characterized by the number of individuals with each strategy and the environmental state, $i=\{Z^m_L$, $Z^m_H, Z^f_L, Z^f_H\}$ and $n$, or $x={i,n}$. Let us write the transitions that increase (decrease) the number of $Z^m_L$ by one as, $T^{mL\pm}_i$, where we include the subscript $x$ denote that the probability of such a transition will depend on the current state of the population and environment.

\begin{align}
\tau T^{mL+}_x &=
    \tau
    ( T^{mH\downarrow mL\uparrow}_x
    + T^{fL\downarrow mL\uparrow}_x
    + T^{fH\downarrow mL\uparrow}_x),\\
\tau T^{mL-}_x &=
    \tau
        (T^{mL\downarrow mH\uparrow}_x
        + T^{mL\downarrow fL\uparrow}_x
        + T^{mL\downarrow fH\uparrow}_x).
\end{align}

For the remaining values we can write the same sets of transitions:

\begin{align}
\tau T^{mH+}_x &=\tau
    ( T^{mL\downarrow mH\uparrow}_x
    + T^{fL\downarrow mH\uparrow}_x
    + T^{fH\downarrow mH\uparrow}_x),\\
\tau T^{mH-}_x &=
    \tau 
        (T^{mH\downarrow mL\uparrow}_x
        + T^{mH\downarrow fL\uparrow}_x
        + T^{mH\downarrow fH\uparrow}_x).
\end{align}

\begin{align}
\tau T^{fL+}_x &=
    \tau
    ( T^{mL\downarrow fL\uparrow}_x
    + T^{mH\downarrow fL\uparrow}_x
    + T^{fH\downarrow fL\uparrow}_x),\\
\tau T^{fL-}_x &=
    \tau 
        (T^{fL\downarrow mL\uparrow}_x
        + T^{fL\downarrow mL\uparrow}_x
        + T^{fL\downarrow fH\uparrow}_x).
\end{align}

\begin{align}
\tau T^{fH+}_x &=
    \tau
    ( T^{mL\downarrow fH\uparrow}_x
    + T^{mH\downarrow fH\uparrow}_x
    + T^{fL\downarrow fH\uparrow}_x),\\
\tau T^{fH-}_x &=
    \tau 
        (T^{fH\downarrow mL\uparrow}_x
        + T^{fH\downarrow mH\uparrow}_x
        + T^{fH\downarrow fL\uparrow}_x).
\end{align}

For larger population sizes, taking an absolute time scale (e.g., the time scale of the resource dynamics), the time between any two updates, $\tau$ is smaller (than at lower population sizes) as the chance that any of the individuals updates increases. Thus, we redefine $\tau\rightarrow \tau/Z$ such that $\tau$ represents the time it takes an individual to reevaluate their decision (instead of the time step between any two updates). Thus, we write the master-equation for the Markov chain defined by $z=i/Z$ and perform a Kramers-Moyal expansion in $1/Z$, keeping the terms of order $1/Z^2$, which, following the approach of Traulsen et al. (2005) results in a Fokker-Plank equation, whose equivalent Langevin equation is 

\begin{align}
\frac{d z^m_L}{dt'}&= T^{mL+}(z,n)-T^{mL-}(z,n)+O(1/Z^{1/2}),\\
\frac{d z^m_H}{dt'}&= T^{mH+}(z,n)-T^{mH-}(z,n)+O(1/Z^{1/2}),\\
\frac{d z^f_L}{dt'}&= T^{fL+}(z,n)-T^{fL-}(z,n)+O(1/Z^{1/2}), \text{and}\\
\frac{d z^f_H}{dt'}&= T^{fH+}(z,n)-T^{fH-}(z,n)+O(1/Z^{1/2}).
\end{align}

Now, we can compute these balances of probabilities in order to get an explicit form of the system of ODE's. 

\begin{align}
T^{mL+}(z,n)-T^{mL-}(z,n) = &
    \;T^{mH\downarrow mL\uparrow}(z,n) - T^{mL\downarrow mH\uparrow}(z,n)\\\nonumber
    &+T^{fL\downarrow mL\uparrow}(z,n) - T^{mL\downarrow fL\uparrow}(z,n)\\\nonumber
    &+T^{fH\downarrow mL\uparrow}(z,n) - T^{mL\downarrow fH\uparrow}(z,n),
\end{align}

which has two types of terms, for intra- and inter-type dynamics. The intra-type dynamics are governed by the terms

\begin{align}
T^{mH\downarrow mL\uparrow}(z,n) - T^{mL\downarrow mH\uparrow}(z,n) &= \frac{Z^m_L Z^m_H}{Z(Z-1)} p(\pi_L-\pi_H) - \frac{Z^m_L Z^m_H}{Z(Z-1)}  p(\pi_H-\pi_L)\\\nonumber
&= \frac{Z^m_L Z^m_H}{Z(Z-1)}\left[\,p(\pi_L-\pi_H)- p(\pi_H-\pi_L)\right]\\\nonumber
&= \frac{Z}{Z-1}z_L^mz_H^m\left[\,p(\pi_L-\pi_H)- p(\pi_H-\pi_L)\right]\\\nonumber
&=\frac{Z}{Z-1}  z^m_L z^m_H \tanh{\left(\beta(\pi_L-\pi_H)\right)},
\end{align}

and the inter-type dynamics are of the form

\begin{align}
    &T^{fL\downarrow mL\uparrow}(z,n) - T^{mL\downarrow fL\uparrow}(z,n)&&\\\nonumber
    +&T^{fH\downarrow mL\uparrow}(z,n) - T^{mL\downarrow fH\uparrow}(z,n)&&
        =\;\frac{Z^m_L Z^f_L}{Z(Z-1)} \hat p(C) - \frac{Z^m_L Z^f_L}{Z(Z-1)} \hat p(-C)\\\nonumber
        &&&\quad+ \frac{Z^m_L Z^f_H}{Z(Z-1)} \hat p(\pi_L-\pi_H+C) - \frac{Z^m_L Z^f_H}{Z(Z-1)}  \hat p(\pi_H-\pi_L-C)\\\nonumber
        &&&\quad= \frac{Z}{Z-1}\left[  z^m_L z^f_L \tanh{\left(\gamma C\right)}  + z^m_L z^f_H  \tanh{\left(\gamma\left(\pi_L - \pi_H + C\right)\right)} \right]. \\\nonumber
\end{align}

If we take the limit as $Z\to\infty$, then consider the limit of weak selection in $\beta$, with $\gamma = \beta\epsilon_2$,  the higher order terms in $Z$ disappear and $\tanh()$ can be approximated linearly, since 

\begin{equation}
    \tanh{(\beta x)} = \beta x + O(x^3) \approx \beta x 
\end{equation}
for small values of $\beta x$.
After re-scaling time so that the $\beta$ terms drop out of the equations, these limits result in the ODE's we considered in the main text, given by 

\begin{equation}
    \frac{d z^m_L}{dt} = z_L^m z_H^m \left(\pi_L-\pi_H\right) + \epsilon_2 z_L^m z_L^f C + \epsilon_2 z_L^m z_H^f \left(\pi_L-\pi_H+C\right). 
\end{equation}
The same arguments applied to the remainder of the strategy dynamical equations also yield the ODE's analyzed in the main text.
\subsection{Environmental dynamics}
We consider a decaying resource that is emitted as a byproduct of strategies deployed by individuals in the population. We will consider a term, $n$ that corresponds to environmental quality. We want to study a system where when all individuals follow the low-impact strategy, the environment approaches its highest state, $n=1$, and when all individuals follow a high-impact strategy, the state of the environment declines to $n=0$. Tilman et al. (2020) show that such a dynamic is mathematically equivalent a re-scaling of the dynamics of pollution emissions. Further, Tilman et al. (2020) show that the dynamics of decaying resources and self-renewing resources are qualitatively equivalent in the context of eco-evolutionary games. Here, we will constrain our analysis to decaying resources since this will lead to fewer model parameters. 

We let $n(t)$ be the state of the environment at time $t$. In a timestep $\tau/Z$, the environment responds to the (constant) current strategy mixture of the population at time $t$ according to 

\begin{equation}
    n(t+\frac{\tau}{Z}) = n(t) + \frac{\tau}{Z}\epsilon_1\left(\frac{Z_L^f(t) + Z_L^m(t)}{Z} - n(t)\right)
\end{equation}
so greater numbers of low-impacts strategists leads to an increasing environmental state, with equilibrium points at $n=1$ when $Z_L^f + Z_L^m = Z$ and $n=0$ when $Z_L^f + Z_L^m = 0 $, as desired. This can be rewritten as 

\begin{equation}
    \frac{n(t+\frac{\tau}{Z})-n(t)}{\frac{\tau}{Z}} = \epsilon_1\left(z_L^f(t) + z_L^m(t) - n(t)\right)
\end{equation}
by rearranging terms and changing from population numbers, $Z_L^m$ to population fractions $z_L^m$. Following the approach of the previous sections where $Z$ gets large while keeping the average time an individual takes to update, we see that the limit as $Z\to\infty$ is of the form

\begin{equation}
    \lim_{\frac{\tau}{Z}\to0}\frac{n(t+\frac{\tau}{Z})-n(t)}{\frac{\tau}{Z}} =\frac{dn}{dt}.
\end{equation}
The equality holds because the expression on the left is the definition of a derivative. This yields an ODE for the dynamics of the environment given by

\begin{equation}
    \frac{dn}{dt} = \epsilon_1\left(z_L^f(t) + z_L^m(t) - n(t)\right)
\end{equation}
which is the same as the dynamical equation considered in the main text. In summary, we have presented a discrete-time individual-based model that converges to the set of ODEs that we analyze in this paper. This representation gives an explicit micro-level motivation and explanation for the structure of the model analyzed in the main text. To get the dynamics of the individual-based model to converge to the system we study, we made several standard assumptions: that the population size is large, that the update-consideration-rate per individual is independent of population size, and that selection is weak.
\section*{Supplementary Information References}
\begin{description}
	\addtolength{\leftmargin}{0.2in}
	\setlength{\itemindent}{-0.2in}
	\item Adamson, M.W. and Hilker, F.M. (2020). Resource-harvester cycles caused by delayed knowledge of the harvested population state can be dampened by harvester forecasting. \textit{Theoretical Ecology}, 13:425-434.
	\item Tilman, A.R., Plotkin, J., and Ak\c{c}ay, E. (2020). Evolutionary games with environmental feedbacks. \textit{Nature Communications}, 11(1):1-11.
	\item Traulsen, A., Claussen, J.C., and Hauert, C. (2005). Coevolutionary dynamics: from finite to infinite populations. \textit{Physical Review Letters}, 95(23):238701.
\end{description}

\end{document}